\documentclass[%
 reprint,
unsortedaddress,
superscriptaddress,
 amsmath,amssymb,
 aps,
prstper,
]{revtex4-1}

\usepackage[explicit]{titlesec}
\titlespacing\section{0pt}{12pt plus 4pt minus 2pt}{3pt plus 2pt minus 2pt}
\titlespacing\subsection{0pt}{12pt plus 4pt minus 2pt}{3pt plus 2pt minus 2pt}
\titlespacing\subsubsection{0pt}{12pt plus 4pt minus 2pt}{3pt plus 2pt minus 2pt}

\usepackage{graphicx}
\usepackage{dcolumn}
\usepackage{bm}
\usepackage{hyperref}

\begin{document}


\title{Preparing for the quantum revolution\\ - what is the role of higher education?}

\author{Michael F. J Fox}
\email{michael.fox@colorado.edu}
\affiliation{%
 JILA, National Institute of Standards and Technology and University of Colorado, Boulder, Colorado 80309, USA\\
 Department of Physics, University of Colorado, 390 UCB, Boulder, Colorado 80309, USA
}%

\author{Benjamin M. Zwickl}%
\affiliation{
 School of Physics and Astronomy, Rochester Institute of Technology, Rochester, New York 14623, USA
}%

\author{H. J. Lewandowski}
\affiliation{%
 JILA, National Institute of Standards and Technology and University of Colorado, Boulder, Colorado 80309, USA\\
 Department of Physics, University of Colorado, 390 UCB, Boulder, Colorado 80309, USA
}%

\date{\today}

\begin{abstract}
Quantum sensing, quantum networking and communication, and quantum computing have attracted significant attention recently, as these quantum technologies could offer significant advantages over existing technologies. In order to accelerate the commercialization of these quantum technologies, the workforce must be equipped with the necessary skills. Through a qualitative study of the quantum industry, in a series of interviews with 21 U.S. companies carried out in Fall 2019, we describe the types of activities being carried out in the quantum industry, profile the types of jobs that exist, and describe the skills valued across the quantum industry, as well as in each type of job. The current routes into the quantum industry are detailed, providing a picture of the current role of higher education in training the quantum workforce. Finally, we present the training and hiring challenges the quantum industry is facing and how higher education may optimize the important role it is currently playing.
\end{abstract}


\maketitle

\section{Introduction}\label{sec:introduction}
The passing of the National Quantum Initiative (NQI) Act~\cite{NQI, stimers2019us, Raymer_2019, monroe2019us} in December 2018 has highlighted the advance of new quantum technologies out of the laboratory and into the commercial environment. These new technologies have the capacity to provide significant advantages to existing industries: from sensing to communication, and, most conspicuously, computing~\cite{pritchard2014uk,DSB2019,NSO-QIS, quantum-supremacy}. The first of the purposes listed in the NQI Act is {``to expand the number of researchers, educators, and students with training in quantum information science and technology to develop a workforce pipeline''}. It is the aim of our research to begin to address this purpose in relation to the role of higher-education institutions. We focus on the training of students (undergraduate and graduate) to enter the workforce and the retraining of the existing workforce. While we do not consider the training of academic researchers (for jobs at universities or national laboratories) or educators, our conclusions may be relevant when considering these groups, especially as the skills needed by the quantum industry are closely aligned with academia. As a result, our goal is to provide a useful resource for faculty and administrative leaders at higher-education institutions who are currently considering how to incorporate the exciting new aspects of quantum technologies into their curricula. Conversely, our goal is not to provide exhaustive quantitative data on companies or employment related to these quantum technologies.

The range of different companies in the industry developing these quantum technologies is large and varied, therefore, to provide some clarity for the ensuing discussion, we first define some key terms we will be using throughout this work. The NQI Act defines quantum information science as  ``the use of the laws of quantum physics for the storage, transmission, manipulation, computing, or measurement of information.''~\cite{NQI}. We use this definition as the basis for our definition of the quantum industry as: \textit{all companies engaged in activities that either apply quantum information science for their product to function or provide technology that enables such a product}. We similarly define the quantum workforce as all the people who work for companies (or specialized divisions within companies) in the quantum industry. We use these definitions in our study to ensure that we do not overlook important parts of the quantum industry.

The results of our study are both compelling and timely because the increased national interest and associated funding opportunities have led higher-education institutions across the U.S. to consider how to provide their students with the skills needed for a career in the quantum industry. Workshops, such as the Kavli Futures Symposium on Achieving a Quantum Smart Workforce (November 4-5$^\mathrm{th}$ 2019)~\cite{Kavliworkshop} and the National Science Foundation (NSF) funded Quantum Information Science and Technology training and workforce development workshop (March 9-10$^\mathrm{th}$ 2020)~\cite{QISTworkshop}, brought together faculty from dozens of physics, engineering, and computer science departments to share how each are developing new courses, certifications, and/or degrees at their institutions. Companies in the quantum industry, being stakeholders in the development of the quantum workforce, sent representatives to these workshops as well. 

The quantum industry has been pro-active in helping with the development of the workforce pipeline. Industry groups, such as the Quantum Economic Development Consortium (QED-C), which was established through the NQI Act with a purpose to support ``the development of a robust quantum information science and technology industry in the United States''~\cite{NQI,QEDC} and the Institute of Electrical and Electronics Engineers (IEEE)~\cite{IEEE}, have established working groups to bring together interested parties to help facilitate education and training related to the skills needed by the quantum industry. As part of these actions the QED-C has conducted a survey of members to quantify the needs of the quantum industry; the IEEE has hosted a Quantum Education Summit at the Rebooting Computing Conference (November 6$^\mathrm{th}$ 2019)~\cite{IEEE-Rebooting}, and will be hosting a Technical Paper Track on Quantum Education and Training at the IEEE Quantum Week (October 12-16$^\mathrm{th}$ 2020)~\cite{IEEE-Quantum-Week}. 

The U.S. Federal Government is promoting interactions between the quantum industry and higher-education institutions through the National Quantum Coordination Office, established by the NQI Act within the White House Office of Science and Technology Policy~\cite{NQI,NQCO-briefing,NQCO-DOE}. Additionally, the NQI Act has directed both the NSF and the Department of Energy to distribute funds. The NSF is doing so by funding 3 Quantum Leap Challenge Institutes~\cite{QLCI} each made up of collaborations between universities and National Laboratories and led by the University of California-Berkeley (quantum computing), the University of Illinois at Urbana-Champaign (quantum architectures and networks), and the University of Colorado Boulder (quantum sensing and distribution). The Department of Energy is providing funding though Quantum Information Science Research Centers~\cite{DOEcenters} led by the following National Laboratories: Argonne (quantum networks, sensors, standards), Brookhaven (quantum computing), Fermilab (quantum materials and systems), Lawrence Berkeley (quantum algorithms and devices), and Oak Ridge (topological quantum materials). These federal-government-led initiatives involve dedicated foci on methods to expand the workforce pipeline for the quantum industry.

In all of the discussions between industry and academia, and within academia, many more questions than answers have arisen. Therefore, before higher-education institutions implement new courses and programs to develop the workforce pipeline, one important question to answer is: \textit{what are the skills that are needed?} We use the term `skills' in the broadest sense: from knowledge of abstract concepts to the ability to build a physical system. To answer this question, we first have to understand what skills are valued by the quantum industry and what the current roles higher education and companies are taking in developing those skills. This follows a similar approach to previous work, in identifying the breadth and depth of skills in the photonics workforce~\cite{Leak2018}. To explore the role of higher-education institutions in the workforce pipeline for the quantum industry, we have conducted interviews with 21 different companies that have self-identified as being within the quantum industry. Each interview covered questions on company context, skills and knowledge required as a function of academic preparation and job type, as well as training and hiring (full interview protocol included in Supplementary Materials Section~\ref{sup:interviewprotocol}). Using this snapshot of the quantum industry, we characterize the existing workforce pipeline by answering the following questions:
\begin{enumerate}
    \item What are the career opportunities that exist in the quantum industry?
    \item What are the skills valued by employers?
    \item How have existing employees gained the required skills?
\end{enumerate}
Once we have established the current state of the quantum workforce, it is then possible to explore how that state may evolve in the future, which can then inform where higher-education institutions can place their efforts to better prepare students who wish to pursue careers in the quantum industry. We do this by presenting data that answers the following questions:
\begin{enumerate}
\setcounter{enumi}{3}
    \item What training and education programs would be helpful to teach the required skills and knowledge?
    \item What are the skills that are currently hard to find when hiring for the quantum workforce?
\end{enumerate}
The answers to these five questions will be provided in Section~\ref{sec:results}. To frame the answers to these questions, we first provide descriptions of the types of activities that are being carried out in the quantum industry.

\subsection{What are the activities of companies in the quantum industry?}\label{sec:typesofcompanies}
The below categories of activities have been developed based on the responses from our interview study, public websites, and the literature on the quantum industry~\cite{NSO-QIS,pritchard2014uk}:
\begin{enumerate}
\item {\bf Quantum sensors:} A company that is developing a sensor, such as a clock, magnetometer, gravimeter, or accelerometer, that has improved precision, compared to existing technology, by taking advantage of the ability to finely control the quantum states of the system, while still being able to be used for commercial applications.
\item {\bf Quantum networking and communication:} A company that is producing quantum-key distribution technologies or software, or  is engaged in the development of hardware technologies to distribute entangled states.
\item {\bf Quantum computing hardware:} A company that is building a quantum computer using any one of many different hardware approaches, such as: superconducting, trapped-ion, or photonic qubits. Additionally, this includes the software development required for the hardware to operate, including, but not necessarily, all the way to a full-stack provision of quantum programming languages to end users who want to run their own quantum algorithms. At the current time, these companies may also be developing software to simulate the operation of a quantum computer on a classical machine. 
\item {\bf Quantum algorithms and applications:} A company that takes a real-world problem and applies knowledge of quantum computation to that problem in an attempt solve it, or at least to demonstrate that it is possible to solve, with the goal of achieving a solution faster than  a classical computer. They may also be involved with the development of new algorithms to run on quantum computers. These are the current `end users' of quantum computing hardware.
\item {\bf Facilitating technologies:} A company who builds, often customized, hardware that is used in either quantum sensors, networking and communication, or computing hardware, such as laser, cryogenic, vacuum, and signal processing components. 
\end{enumerate}
We use the above terms, when referring to companies, in order to protect the identity of the companies that participated in the research. As companies may have activities in more than one of the above categories, there are subtleties that must be considered even when asking what may appear to be a simple question, such as: ``how many quantum computing companies are there?'' We explore some of these subtleties relating to our data below.

\subsection{Distribution of company types}
\begin{figure*}
    \centering
    \includegraphics[width=\textwidth]{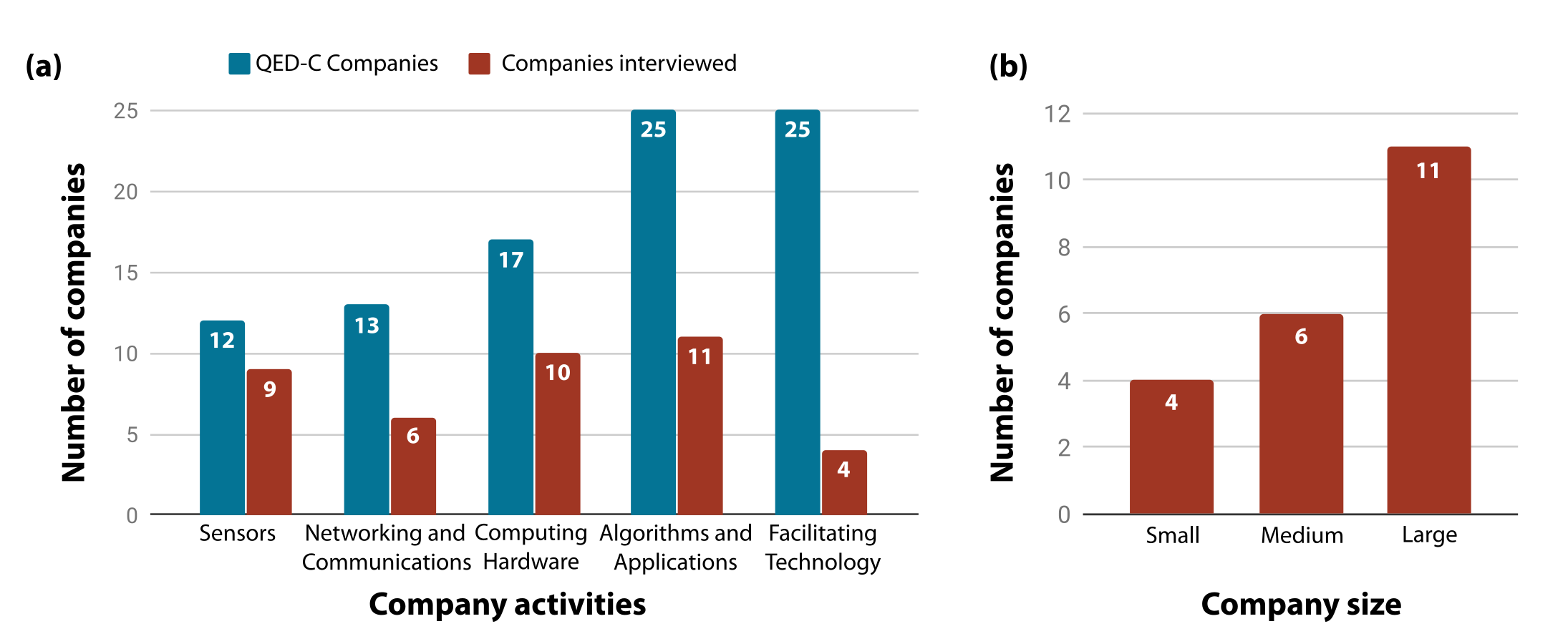}
    \caption{Distribution of companies in the U.S. quantum industry. (a) The number of companies with letters of intent signed with the QED-C as of February 2020 (blue) and the number of companies interviewed (red) plotted against the types of activities. Each individual company may be involved in more than one activity, so the blue bars do not sum to equal the 87 companies in the QED-C, nor do the red bars sum to equal the 21 companies interviewed. 20 companies in the QED-C were not categorized due to lack of information about their activities. (b) The distribution of the 21 companies interviewed based on the total number of employees in the company: Small is less than 20; Medium is between 20 and 200 inclusive; Large is greater than 200. The corresponding information for QED-C companies was harder to assess reliably, and, therefore, has not been included (most of the presented numbers were ascertained during the interviews).}
    \label{fig:TypesAndSizes}
\end{figure*}

To describe the landscape of the quantum industry in terms of the distribution of companies among the types of activities described previously, we researched the companies that have signed letters of intent with the QED-C~\cite{QEDC} and applied our categorization to those companies. While not all companies that have activities within the quantum industry have joined the QED-C, it is nevertheless informative to discuss what the distribution of companies reveals and how that relates to the distribution of companies within the sample interviewed for this research (Fig.~\ref{fig:TypesAndSizes}(a)).

The type of activity that corresponds to the largest number of companies in the QED-C is that of facilitating technologies. Based on data from our interviews and public information, these companies are often small (less than 20 employees) to medium (between 20 and 200 employees inclusive) sized, in terms of total number of employees (c.f. Fig.~\ref{fig:TypesAndSizes}(b)), that specialize in components (e.g. lasers) that may be used in other industries, as well as the quantum industry. Some of these companies are manufacturers of components that have uses in many different industries, while others have experience providing specialized equipment to university research laboratories. 

The other activity with the largest number of companies in the QED-C is quantum algorithms and applications. This number includes almost all of the quantum computing hardware companies. We make the distinction between these two types of activities because both the product and the skills required of employees are very different. Nevertheless, these two types of activities are clearly related, and are often both described as `quantum computing'. The companies with activities based solely on the applications of quantum computing are generally small companies that provide consulting on the possibilities of quantum computation to larger companies who are themselves not actively involved in the quantum industry. Of the quantum algorithms and applications companies interviewed, not all were also quantum computing hardware companies. These small quantum algorithms and applications companies can exist because of the ability to remote access the hardware provided by the medium-to-large sized quantum computing hardware companies (large being greater than 200 employees). This means that these small companies do not have to overcome the significant barrier to entry into the marketplace of building and maintaining quantum computing hardware.

Quantum networking and communication companies are mostly medium-to-large in terms of number of employees. In 2017, it was reported that ``U.S. interest in QKD [quantum-key distribution] has declined''~\cite{IDA2017}, however, partly due to the NQI, this is an area that has  been recently reinvigorated~\cite{NQCO-Quantum-Networks}. 

Our sample indicates that quantum sensing companies either are small, in terms of number of employees, or are small sub-divisions of large companies. They often use similar technologies to some of the approaches to quantum computing. Quantum sensing is the most established activity within the quantum industry (ignoring facilitating technologies), as the atomic clock is a quantum sensor that has been commercially available since 1956~\cite{forman1985atomichron}. The improved precision offered by recent developments in hardware and quantum information theory are currently making their way out of the laboratory and into new commercial devices that benefit from reductions in size, weight, and power~\cite{DSB2019}.

In the following section, we present the methodological approach we used for this study and discuss some of the limitations that may apply to our conclusions. In Section~\ref{sec:results}, we present the results of our interview study comprising 21 companies that span the full range of the types of activities within the quantum industry. In Section~\ref{sec:discussion}, we discuss the results and present our conclusions on what the role of higher education is in preparing for the quantum revolution.

\section{Methodology}\label{sec:methods}

Initial research questions were developed to explore the needs of the quantum industry. These were translated into an interview protocol (Supplementary Material Section~\ref{sup:interviewprotocol}), that was designed to last approximately one hour. The interview was semi-structured, meaning that deviations from the script were allowed for clarification of statements from the interviewee, as well as to ensure a complete coverage of the interview material. The interview was tested on a colleague before being administered to attain feedback on logical self-consistency, wording of questions, and coverage of topics. Minor alterations to the interview protocol were incorporated between interviews in order to improve the clarity of the questions. The lists of scientific and technical skills used as prompts in the interview protocol were taken from the 2016 report of the Joint Task Force on Undergraduate Physics Programs~\cite{Jtupp2016}.

Companies from the U.S. engaged in activities that fall within the quantum industry were contacted through the QED-C (approximately 70 companies at that time) on September 22$^\mathrm{nd}$ 2019. Specifically, persons who were involved with the hiring and supervision of new employees were asked to contribute to the study. The companies were provided with anonymity. Seventeen companies responded to the request, leading to 11 interviews. The sampling of companies was expanded beyond the initial responses to the QED-C request through snowball sampling~\cite{Merton1948,Biernacki1981} and the direction of the sampling was chosen to sample across the quantum industry (c.f. Fig.~\ref{fig:TypesAndSizes}). In addition to the QED-C mailing list, emails were sent to 22 individuals to solicit participation, which led to 11 further interviews. In some cases, we interviewed more than one person at the same company in the same interview (the maximum number of interviewees at one time was 3). Therefore, we spoke with 26 individuals in a total of 22 interviews. Two of these interviews were with representatives from the same company, therefore resulting in 21 companies in our sample. The interviews were conducted in person or via teleconferencing. The companies were categorized based on their activities as defined in Section~\ref{sec:typesofcompanies}. The definitions of the categories for the companies were revised based on the information gained from the interviews.

The interviews were transcribed using an automated, online transcription service (Rev.com). Each automated transcription was checked against the original audio recording and transcription errors corrected. An \textit{a priori} codebook was developed based on our research questions~\cite{saldana2015coding}. The key codes included job types, degree level, and degree subject mainly to act as labels to associate with identified skills. Emergent codes, mainly identifying the distinct skills and knowledge mentioned in the interviews, were also established on the first coding pass of the transcripts and added to the codebook~\cite{saldana2015coding,otero2009getting}. After an initial coding of 8 transcripts, confirmatory inter-rater reliability (IRR) was completed by an independent coder on 1 transcript, whereby the independent coder would state whether they agreed or disagreed with the assignment of codes by the original coder. This led to refinements of the codebook and re-coding of the 8 transcripts, before coding the remaining transcripts. Confirmatory IRR only checks for false-positive coding, which is useful for ensuring the clarity of the codebook and the applications of the codes. 

A second round of IRR was completed once all transcripts were coded. For this second test of IRR, a subset of the transcripts coded as either a job type, degree subject, or degree level was sent to a second independent coder to apply the same codes (without being aware of the original code assignments - i.e. this is not confirmatory IRR). We report, in Table~\ref{tab:IRR}, the average initial agreement between the two coders, at sentence-level matching, using two measures. These high levels of agreement demonstrate the robustness of the analysis, and indicate the level of uncertainty in the numerical results~\cite{Hammer2014}. Additionally, confirmatory IRR was completed on the skills and knowledge codes, with an average agreement of 98\%. All the disagreements were discussed and reconciled.

\begin{table}[t]
\centering
\caption{\textbf{Results from inter-rater reliability comparison of coding}. Agreement is the percentage of the sample of the interview transcripts that both coder A and coder B applied the same code to plus the percentage of the sample to which neither coder A and coder B applied that same code. Cohen's Kappa is a statistical measure of agreement accounting for the possibility of chance agreement between two coders~\cite{Cohen1960}. The maximum value of kappa is 1, implying complete agreement. A value of 0.74 is generally considered to be a sign of good agreement~\cite{Byrt1996}.  }\label{tab:IRR}
\begin{tabular}{r|cc}
               & Agreement/\% & Cohen's kappa \\\hline\hline
Job types      & 98           & 0.74          \\
Degree level   & 98           & 0.78          \\
Degree subject & 95           & 0.74         
\end{tabular}
\end{table}
\vspace{-5pt}

\subsection{Limitations of the research}
There are a number of limitations to the results that will be presented, some are dependent on the methodology used for collecting and analyzing the data, while others are due to the nature of the subject being studied. Of the former, the qualitative nature of the investigation means that, while we gain an understanding of the depth and variation of the types of skills and knowledge required across the quantum industry, we are unable to report, for example, how many software engineers are employed currently, nor make numerical estimates for future employment numbers. However, we emphasize that this is also one of the advantages of this study, as companies were more willing to contribute given we were not asking about strategic information. A second methodological limitation is in the sampling of companies. We relied on the QED-C for initial contact with companies, then used information from interviews to identify further companies and individuals to contact. While the initial responses from the QED-C led to half of the interviews, removing selection bias on the side of the authors, the snowball sampling introduces bias based on already existing connections between individuals. Additionally, no more than 3 people were interviewed from any one company, which means the interview may not be representative of the whole company, especially if the company is engaged with different aspects of quantum technologies, but the interviewee has experience limited to one area, e.g. hardware. Finally, given the background of the authors, it is possible that there was bias towards physics in the identification of interviewees, as potential interviewees who were themselves physicists might have been more receptive to talking to researchers based in a physics department. However, this is also an advantage, as interviewees were able to describe in depth the technical details of the skills and knowledge required by their employees. 

The analysis of the data depends on the application of the coding scheme, the consistency of which has been tested through IRR. The numerical analysis resulting from this, nevertheless, has limitations, especially when investigating relationships between codes. This is because the codebook allows for some codes to be applied only where there is an explicit reference to the item being coded, or that it is clearly implied from the context of the surrounding discussion. For example, to apply the code of the degree level of ``Ph.D.'' would require the interviewee to make an explicit statement about employees with a Ph.D. Codes were similarly applied for both degree level and job type. Therefore, while a large time might have been spent in an interview discussing Ph.D. physicists who were employed as experimental scientists, this would only have been coded with all three codes when the degree level (Ph.D.), degree subject (physics), and job type (experimental scientist) occurred explicitly in the same context. Therefore, analysis, such as that presented in Section~\ref{sec:employees-gained-skills}, only represent a subset of the data that has been collected. 

Of the limitations due to the nature of the subject being studied, the first is that the companies responding from the QED-C self-identified as being in the quantum industry and were interested in the development of the workforce. Therefore, companies who did not consider themselves to be in the quantum industry would not have, at least initially, been sampled. This is especially relevant when considering the role of facilitating technology companies, as unless these companies are providing bespoke solutions to quantum companies, as the ones who are included in this study were, their connection with the products of the quantum industry may be limited. This is related to the wider question of which employees are in the quantum industry: is it just the ones who need some knowledge of quantum physics or quantum information, or is it everyone in a company that designs, makes, sells, and supports such products? How this question is answered internally by a company strongly relates to whether they would have contributed to this study. The companies that contributed to this study had sufficient prior engagement with the premise of the study to take part, and so this introduces some bias based on the construction of the current view of the quantum industry towards companies with that capacity, however that bias is not unique to this study. Finally, as we have already discussed, there is no agreed upon definition of what the quantum industry is, which is in part due to the rapidly evolving nature of the industry. This means that there is a temporal limitation to the validity of this study; as the technologies advance different skills will be needed to those documented here and as higher-education institutions augment their degrees and courses, the supply-side of the workforce will also change. The hope being that those two aspects align and complement each other. Therefore, this study can be classed only as a snapshot of the quantum industry in the Fall of 2019.


\section{Results}\label{sec:results}

We now provide our answers to our research questions posed in Section~\ref{sec:introduction}. The numerical results presented below are based on the number of companies to which each given code was assigned (e.g., part of a transcript being labeled with a specific skill), rather than the total number of occurrences of a code across all interviews. These numbers are representative of our sample and not necessarily the general population of companies in the quantum industry. Additionally, while the numbers do indicate the frequency of responses to our interview questions, frequency is not always equivalent to importance.

\subsection{What are the career opportunities that exist in the quantum industry?}\label{sec:career-opportunities}
\begin{figure}[t]
    \centering
    \includegraphics[width=\linewidth]{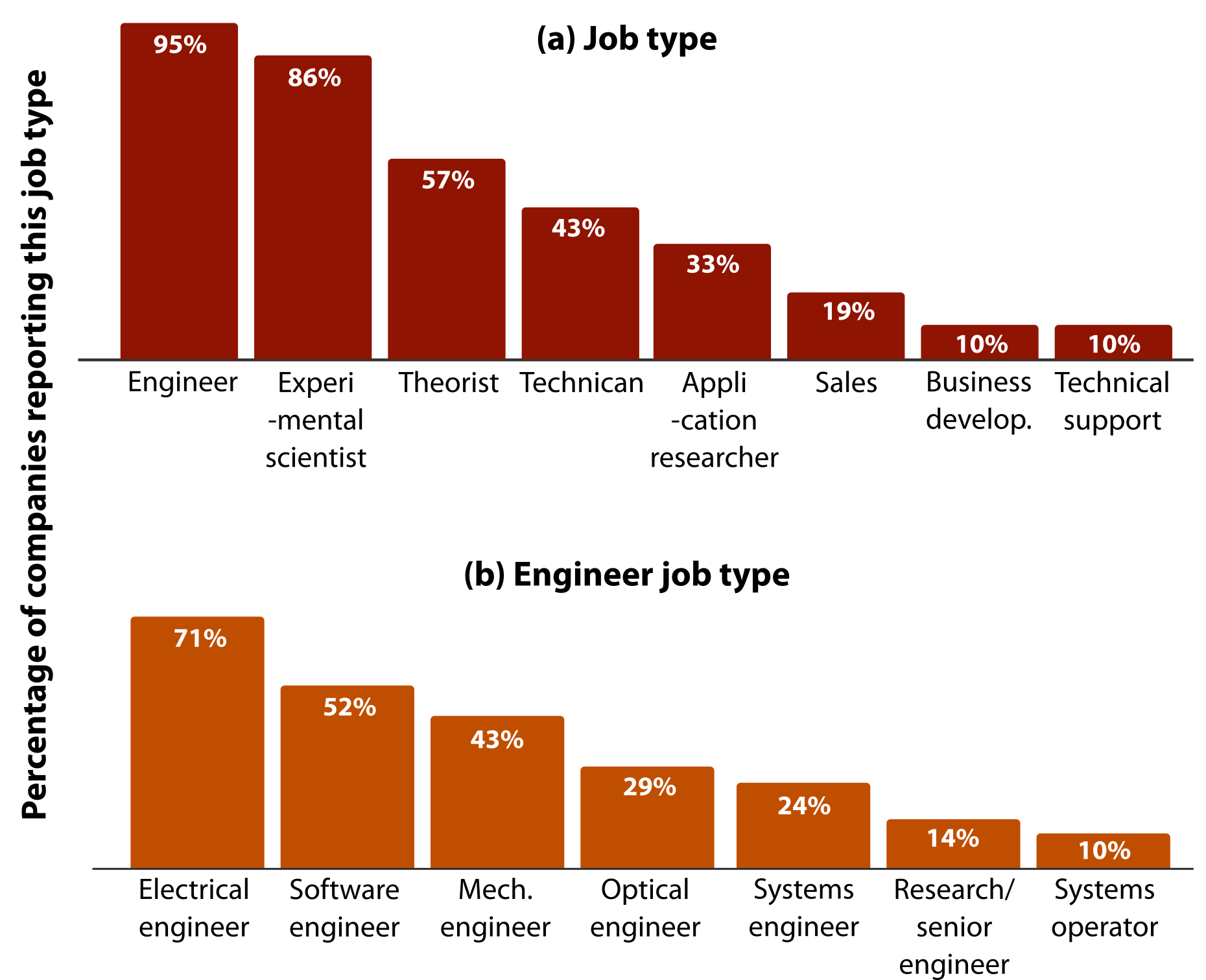}
    \caption{Jobs within the quantum industry. (a) For all major job types. (b) For only job types identified as engineering. In both (a) and (b), each bar represents the percentage of the 21 companies interviewed that indicated they have employees in the types of jobs labeled. Jobs that were identified by only one company are not included in the analysis to avoid identification of that company.}
    \label{fig:JobTypes}
    \vspace{-5pt}
\end{figure}

We have identified five main types of technical careers available within the quantum industry: engineer, experimental scientist, theorist, technician, and application researcher (Fig.~\ref{fig:JobTypes}(a)). Almost all the companies interviewed reported that they had job positions titled ``engineer''. To understand the role of an engineer in the quantum industry, we have further classified those engineering jobs within the different sub-disciplines of engineering (Fig.~\ref{fig:JobTypes}(b)). Later, in Section~\ref{sec:discussion}, we raise the question of whether these are quantum engineers. We emphasize that this is our own categorization of job positions in companies, and the titles we have associated with each job are not always the same as those in any given company. Furthermore, these job titles should not be confused with the degrees that employees may have earned, which we discuss later. There is also some overlap between these different jobs depending on the company, for example someone employed as an experimental scientist may need to have a greater experience of skills normally associated with an engineer when working for one company compared to another.

In the Supplementary Materials (Section~\ref{sup:skills-by-job}), we provide rich descriptions of each of these jobs and what skills are associated with those roles. In these descriptions, the blurred lines between the academic disciplines of physics and engineering and the skills needed in the workplace become evident. An interesting, and outstanding, question is: how are the jobs described distributed within a company? This distribution is a sensitive detail for some companies, as it can reveal what stage a company is at in their product development with later stages having a higher proportion of engineers and technicians in their workforce to enable manufacturing. We deliberately did not collect these data in order to encourage participation in this study, however, in the following section, we identify which skills are most often cited by companies as important for their employees to have. Some of these skills transcend individual jobs described above, while others are more specific.

\subsection{What are the skills valued by employers?}

A deep knowledge of the theory behind quantum information science is not a necessary or sufficient requirement to work in the quantum industry. Indeed, `classical' skills are highly valued. We have already seen this in our discussion of the variety of different types of jobs within the quantum industry. 

\subsubsection{Shared skills across the quantum industry}

We now consider the skills and knowledge valued in the quantum industry independent of the specific job types and company activities. Our goal is to create a resource to which faculty may refer to when teaching existing, or developing new, courses. First we consider: what skills are valued by companies across the quantum industry? Remembering that not all employees need all of these skills.

Almost all the companies mentioned the importance of coding skills (90\%) and experience with using statistical methods for data analysis (90\%). Coding skills are needed for the design and control of experimental apparatus, as well as the collection and analysis of the data from that apparatus. Coding skills are also needed for the collaborative development of software environments through which a user may interact with the hardware. Data analysis is required at both the fundamental, analog level of signal inputs and outputs from a piece of hardware, and at the abstracted level, such as processing the output from a quantum system (repeated sampling from a probability distribution) and interpreting its meaning.

Coding and data analysis skills are related to the expectations of most companies that employees would have laboratory experience (81\%), which indicates that hardware development is a key component of the quantum industry. Only pure quantum algorithm and application companies do not need any experience in a laboratory. Experimental scientists, with a Ph.D., would ``have experience starting an experiment in their lab and know what it takes to get something up and running.'' For junior employees, with a bachelor's or master's degree, a senior design/capstone project in a quantum lab, or a similar internship, is a major plus. An essential aspect of laboratory experience is gained from teaching laboratories, where it is expected that students have learned: ``how to keep a lab book ... how to document what [they've] done... how to prepare a report... how to propose a hypothesis.''

Having knowledge and experience with electronics is almost as ubiquitous an expectation as coding (76\%). Electronics is used to control and power the hardware (lasers, microwave antennae, etc.) used to manipulate and measure the system storing the quantum information. Some of this electronics is standard control systems, while other pieces need to be manufactured to distinct specifications for the system, often with low-noise requirements. 
Troubleshooting and problem solving are also valued by companies (71\%). These skills are related to both experience in a laboratory environment and debugging computer code. Companies recognize that it is hard to assess whether someone is good at troubleshooting or problem solving, which is why they value seeing practical experience on a potential employee's resume.

Material science and knowledge of material properties relevant to a company's specific hardware play an important role when designing and building new hardware, and so many companies (67\%) look for that knowledge in their employees, e.g.: ``material specialists [who have] actually built the superconducting circuits''. While some companies simply need this knowledge in their engineers and experimental scientists to design hardware, other companies have teams that actively develop new materials. This latter group would require more in-depth knowledge of material science; needing to have a good background in quantum mechanics and condensed matter systems, but not necessarily experience in quantum information science.

The most valued skill related to quantum information science is knowledge of quantum algorithms and computer science (62\%). This category is almost exclusively related to quantum computing companies, though some algorithms for quantum information processing are utilized in sensor and communication (cryptography) activities. There are a number of different aspects of this skill: 1) development of new algorithms; 2) implementation of existing algorithms on hardware; and 3) application of existing algorithms to specific problems. For the first of these, a deep knowledge of computer science and mathematics is required, though little physics is necessary. The second requires more physics knowledge, as it relates to how to translate from the abstract space of operations on single or pairs of qubits to real microwave, radio frequency, or laser pulses to perform those operations, accounting for real world effects. The third assumes the previous two aspects exist, and so requires relatively less knowledge of these (except when debugging). Instead, value comes from knowledge of the set of algorithms that exist, what they are useful for, and how to run them on a simulated or real device.

\subsubsection{Variation in skills required by the quantum industry}
\begin{table*}[t]
    \centering
    \caption{The types of skills and knowledge that may be relevant for possible a course on `real-world quantum information theory'. The grayed region indicates skills that are shared across multiple courses (see Section~\ref{sec:skillstables} for tables for other courses). The examples given are quotes from our interview study and are provided to give context to how the skills are useful in the quantum industry.}
    \includegraphics[width=0.9\textwidth]{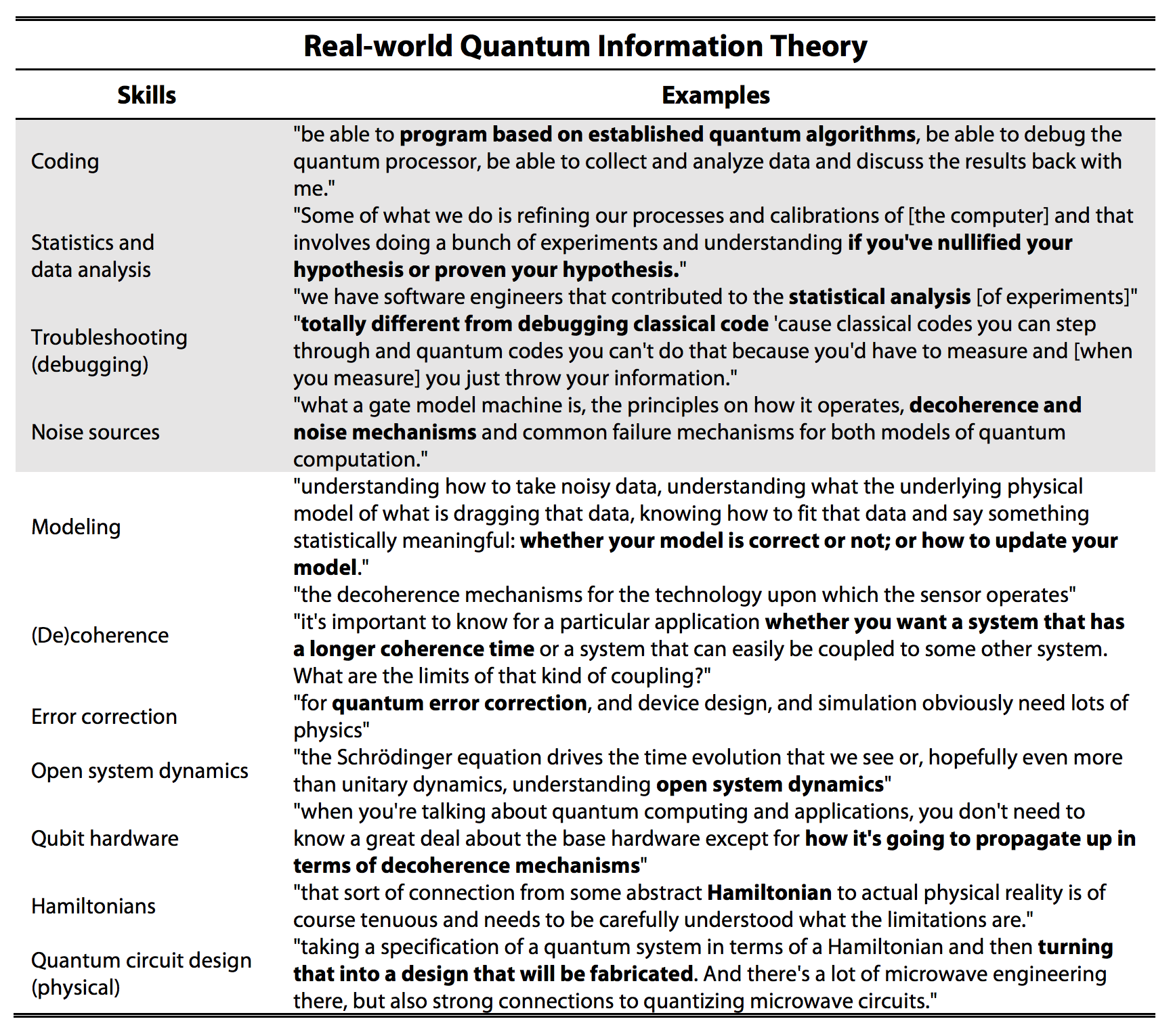}
    \label{fig:Real-worldQIS}
\end{table*}

Beyond the skills discussed above that are shared across multiple activities in the quantum industry, many skills become very dependent on the specific technology being developed. We have chosen to categorize these skills by taking the view of a faculty member wishing to identify how their new or existing course material is relevant to the quantum industry. We categorize the skills into what may be considered titles of potential courses in a quantum information science and technology curriculum: Traditional quantum theory; Quantum information theory; Real-world quantum information theory; Hardware for quantum information; Electronics; Mechanical engineering; Optics and opto-mechanics. We have ordered the courses from theoretical to experimental. These groupings of skills emerged from the coding analysis of the interview transcripts (see Section~\ref{sec:methods}), whereby one of the authors and an independent coder separately categorized the skills that had been coded. The names for each potential course were assigned after the skills were grouped. Comparison of these two categorization schemes showed good agreement and led to minor refinements of the categories. For each skill, we have provided at least one quote from our interview study to provide insight into how that skill is relevant to a particular area of the quantum industry. These quotes are exemplary and should not be considered prescriptive in their implications for course design.   

Tables presenting the categorization of the skills can be found in the Supplementary Materials  (Tables~\ref{tab:trad-q-theory}-\ref{tab:optics}). We envision two uses for these by instructors: either to provide inspiration for topics to cover when developing a new course, or to look up topics that an instructor may already be teaching for relevant examples that can be related to applications in the quantum industry. To illustrate our categorization and these potential uses, we include here one of the tables (Table~\ref{fig:Real-worldQIS}) for the categorization of possible components of a course in real-world quantum information theory. In this example, we see that an instructor developing a new course may choose to focus on statistical data analysis and how to deal, from a theoretical perspective, with the practical issues of noise affecting the results of quantum computation. Different institutions and instructors may want to focus on different skills included in Table~\ref{fig:Real-worldQIS}, depending on their own priorities and those of their students, such as emphasizing the noise issues in one specific type of qubit hardware. For an instructor of an existing quantum course, who wishes to include a modern context or application, they may identify in Table~\ref{fig:Real-worldQIS} that they can relate their existing content to problems in the quantum industry through how the time evolution defined by the Schr\"{o}dinger equation is important in determining the open-system dynamics of qubits and hence their decoherence.

\subsection{How have existing employees gained the required skills?}\label{sec:employees-gained-skills}
\begin{figure*}
    \centering
    \includegraphics[width=0.8\textwidth]{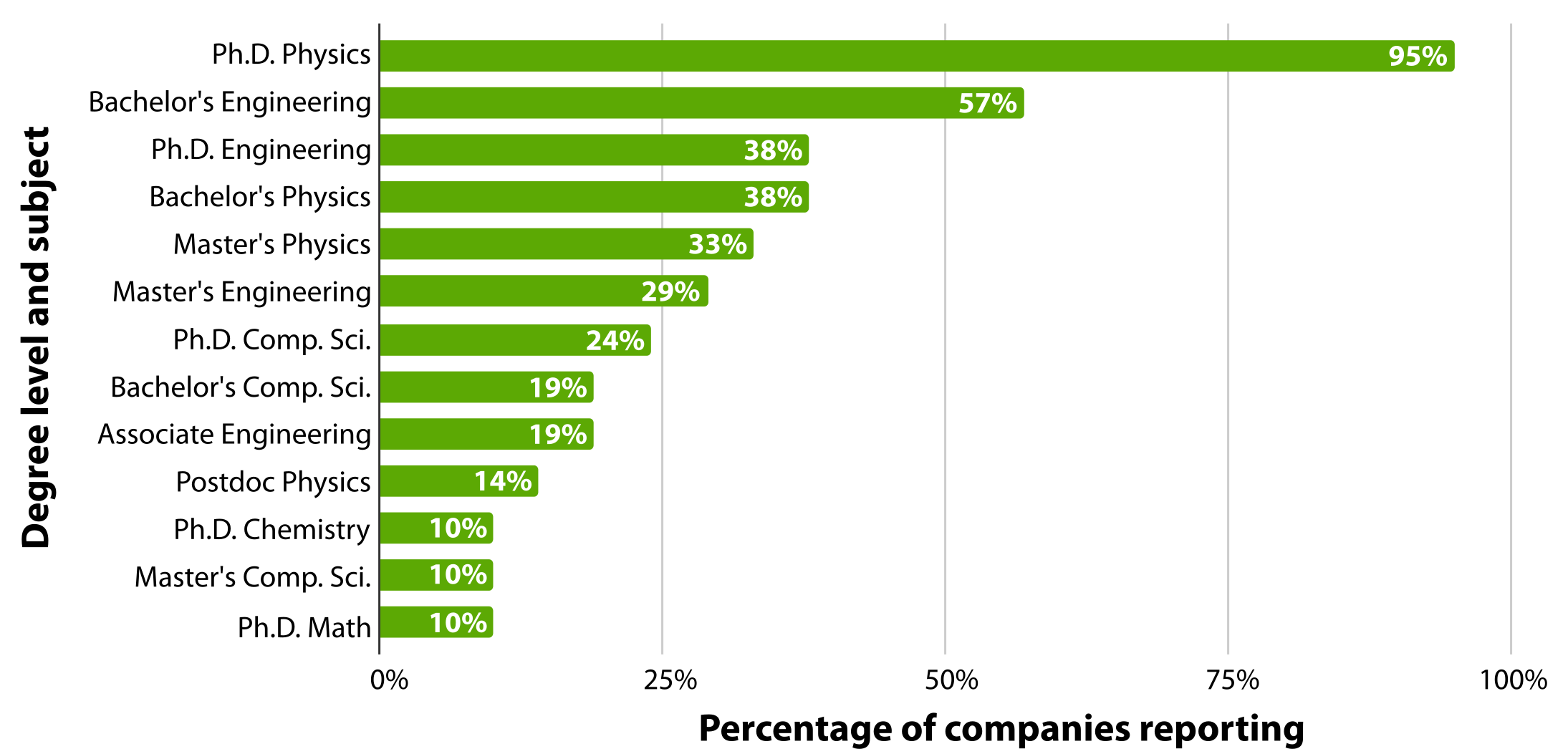}
    \caption{The top 13 degree and subject combinations found in the quantum industry. The percentage corresponds to the number of companies, of the 21 in our sample, reporting at least one employee with the given combination of degree and subject.}
    \label{fig:OnlySubjectLevel}
\end{figure*}
Existing employees in the quantum industry have gained the skills discussed above through both higher-education courses and learning through opportunities provided on the job. In this section, we aim to document the current situation where skills are acquired in order to provide the background for later recommendations. We first present data on the routes through academia that employees have taken, and then detail how companies are providing further training to their employees.

\subsubsection{Routes into the quantum industry through higher education}
\begin{figure*}
    \centering
    \includegraphics[width=0.8\textwidth]{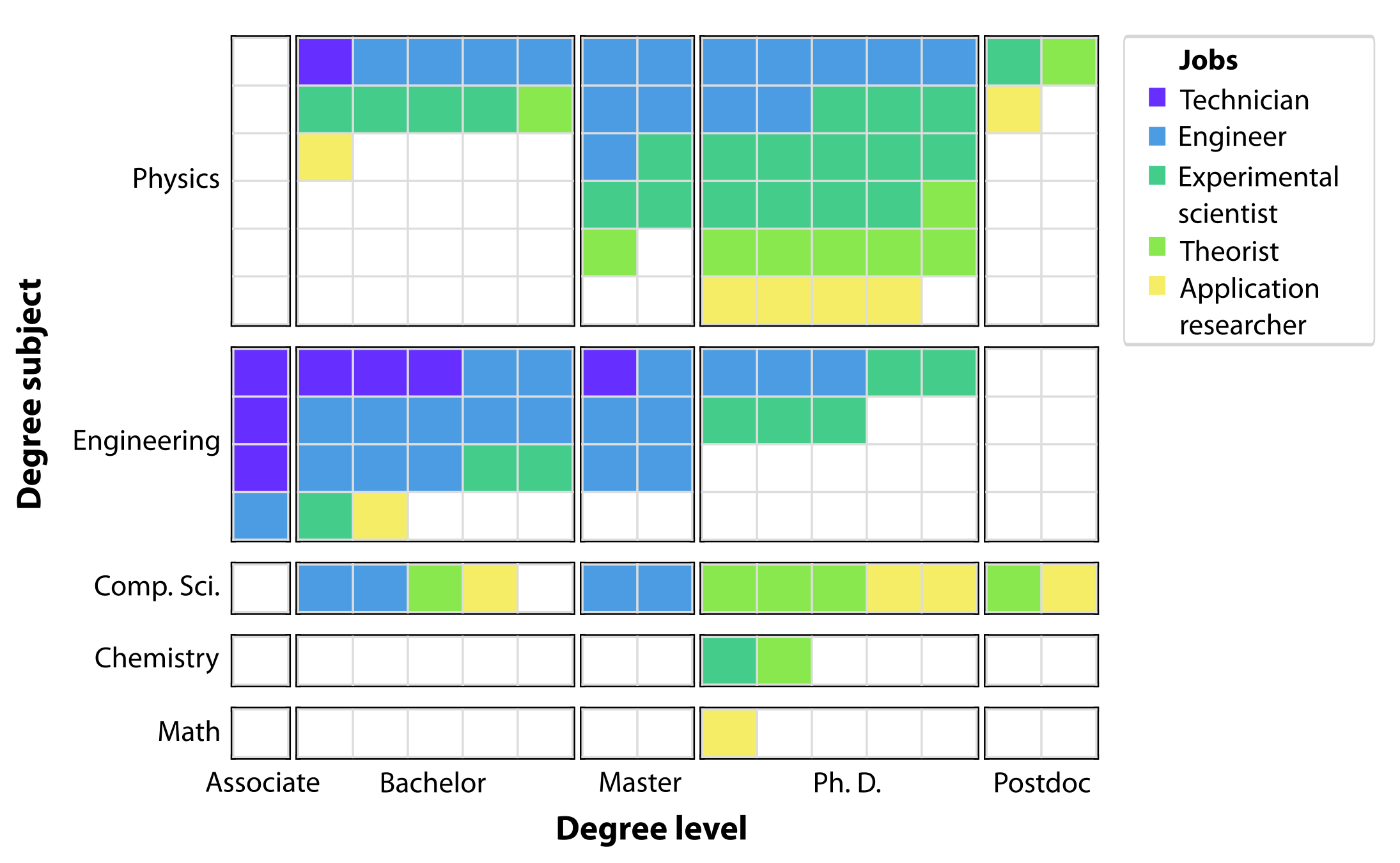}
    \caption{Jobs by degree level and degree subject. Each colored box represents the response of one company where they indicated at least one existing employee occupied the given job (denoted by the color of the box) and had studied the given subject to the degree level shown. The number of rows and columns assigned to each subject and degree is an aesthetic choice. Note, this figure does not represent the number of employees with each job; Fig.~\ref{fig:OnlySubjectLevel} gives a better representation of the distribution of employees by degree level and subject. For a discussion of the limitations to the numbers in this plot see Section~\ref{sec:methods}.}
    \label{fig:DegreeSubjectLevel}
\end{figure*}

Our data indicate that higher education provides routes into the quantum industry through the traditional disciplines of physics, engineering, computer science, math, and chemistry (Fig.~\ref{fig:OnlySubjectLevel}). In the Supplementary Materials (Section~\ref{sup:skills-by-degree}), we provide detailed descriptions of how each degree level and subject prepares a student with the skills necessary to work in the quantum industry. The majority of companies reported having at least one employee with a Ph.D. in Physics. This seems at odds with the results presented in Fig.~\ref{fig:JobTypes}(a), where the majority job type was an engineer, however, while employees may be trained in physics, they move into engineering roles when they join a company - again raising the question: what is a quantum engineer? We can see this clearly in Fig.~\ref{fig:DegreeSubjectLevel}, which relates the degree subject and level to the different job types. In Figure ~\ref{fig:DegreeSubjectLevel}, Ph.D. physicists hold jobs of all types, except for the technician role. Furthermore, we see that fewer companies have employees with engineering Ph.D.s than physics Ph.D.s, but employ more bachelor's degree engineers than bachelor's degree physicists. This reflects the skills and experience that the company is looking for when they are hiring for specific jobs, details of which can be found in the Supplementary Materials.

\subsubsection{Training provided by companies}\label{sec:training-provided-by-companies}
There is a general recognition that new graduates are almost never ready to seamlessly start to work in the quantum industry, ``just because someone has a Ph.D. doesn't mean they're ready to work''. It is inevitable that numerous skills have to be learned while employed. We present the skills identified by more than one company where training has been required, as a function of the different types of job. However, this results in very few identifiable skills, due to the fact that a lot of the training that takes place within each company is very specific to the products that they are developing. This domain-specific knowledge can be learned only while in employment, as it includes not only proprietary information about product design, manufacture, and operation, but also knowledge about how the company functions. This puts an upper limit on the extent to which higher education can prepare students to enter the workforce. We must acknowledge that companies will always play an essential and complementary role to higher education.

Most companies do not have a formal structure to their training, expecting new employees to learn ``on the job'' (95\% of companies interviewed). This informal education comes mainly through independent learning (76\%), i.e., ``pick up a textbook, read some papers'', or through online courses or tutorials (33\%). This independent learning is often guided by senior or peer mentors (62\%) and is personalized to the needs of each employee and their role. Employees are also expected to learn from internal seminars and group meetings (14\%). In the Supplementary Materials (Section~\ref{sup:training-provided-by-companies}), we provide a breakdown of the on-the-job training provided by companies for each of the 5 main job types. We discuss more structured forms of training in the following section, when describing the specific training needs related to each degree level.



\subsection{What training and education programs would be helpful to teach the required skills and knowledge?}
In the preceding sections, we have presented data that help us evaluate the current role of higher education in training employees for careers in the quantum industry. Now, we present data from asking employers how higher-education institutions could better prepare students for careers in the quantum industry. We emphasize that the results presented here are those from the sample of companies interviewed and therefore do not take into account the views of teaching or research faculty who might have different priorities when educating students. 

The training provided by Ph.D. programs is generally seen as the best preparation to enter the quantum industry, ``we're still at a point where we usually need a Ph.D. for them to be useful.'' Standard graduate physics courses taken during the first years of a Ph.D. are seen as adequate preparation (``what I expect is a standard, graduate-level physics knowledge,'' including: quantum mechanics, electromagnetism, atomic physics, statistical mechanics). In addition to the specific domain knowledge gained over the years of a Ph.D., the experience of doing research and developing one's own project are key strengths of completing a Ph.D. The only changes that were mentioned included: 
\begin{enumerate}
    \item more experience with software development in a collaborative environment (using tools such as Git);
    \item team-working skills;
    \item engineering and system design skills; and
    \item more awareness of how business works.
\end{enumerate}
The first two of these are valuable in academic careers as well as industry, while the latter two could be harder to include in Ph.D. programs, and could justifiably be the purview of industry training. Notably, these changes would help many students looking for employment even outside of the quantum industry.

As we have seen above, companies are hiring bachelor's and master's level engineers, physicists, and computer scientists, with the expectation that these employees will be trained by the companies. There is one thing that the quantum industry requested from higher-education institutions to help in this transition, which is a one or two semester course in order to increase quantum awareness (33\% of all companies interviewed): ``a basic course in quantum information for engineers and I guess there could be sort of two different versions of this... one is the more sort of algorithms and applications and software and programming languages one and the other one is more like device physics and qubits and error correction and control electronics kind of one.'' The hardware track would be geared towards electrical, mechanical, and optical engineering students, while the quantum software track might be more focused on software engineering and computer science students. By recruiting more from a diverse range of degree subjects and levels than the currently dominant physics Ph.D. programs, there is a larger pool of possible employees. This approach has been independently recommended by the Defense Science Board that the Military Department Academies ``should add a one-semester quantum technology class for engineering, science, and computer scientists''~\cite{DSB2019}. Quantum awareness has also been highlighted in the National Strategic Overview for Quantum Information Science~\cite{NSO-QIS}. Furthermore, this trend fits with the growth of the companies as they move their products out of development into production and the ratio of engineers to physicists increases - the technology is ``being transitioned to a product and so at that point we start wanting to pull in more engineers, more technicians.'' This change occurs as the science problems are solved, and the main issue becomes ensuring the system is reliable, making classical engineering skills become even more valuable.

A high value is associated with research experience, as evidenced by the large number of Ph.D.-holding employees. Companies like to see this experience in bachelor's and master's students where they have taken internships (19\%) or worked in a research laboratory (48\% of companies reported hosting interns - mainly graduate students - and use the internship to help their own recruitment). There is a desire for engineers to have the opportunity to do quantum related projects in senior design/capstone projects: ``And they went and worked in labs and built low-noise laser controllers or high-ish frequency RF stuff.'' A few companies interviewed (14\%) indicated that a master's in quantum information science, with laboratory projects or internships, would be a good way to prepare students who had not gained that experience in their undergraduate program, but who also did not want to do a Ph.D. before entering industry.

The other area of value that higher education can add to the quantum industry is in retraining existing employees. This is beyond employees enrolling in existing courses. Companies report a desire for on-demand short courses of no more than a couple of weeks, or a few hours per week over a longer period of time (52\%), tailored to specific technologies: ``day long workshops and courses... like [an] optical-metrology workshop or vacuum for dummies''; ``a short course in cryogenics, or... a very short course on microwave electronics.'' There is some interest in providing courses to increase general quantum awareness for existing employees, though the value of that to a company, where this could be provided internally, is not very high: ``I wouldn't care to necessarily send our engineers off to a quantum computing mini-course, except for the fact that that's interesting and enriching. It's of less value.'' The content most valuable for companies is through gaining hands-on experience with new laboratory technology (33\%), such as a one-week course where ``we actually went there and built a [frequency] comb'', or that employees ``can now align MOTs [Magneto-Optical Traps] or laser systems''. This is because there is a high cost associated with purchasing and setting up equipment, as well as a risk associated with it not working, that can be mediated by making use of resources that already exist at university research facilities. The exact content of such short courses would depend upon the available expertise at a higher-education institution as well as industry demand.

\subsection{What are the skills that are currently hard to find when hiring for the quantum workforce?}
The challenges of hiring are not simply a function of the number of new graduates from higher-education programs, but also depend on the demand for skills across the quantum industry, as well as other industries, and the situation of each company in terms of its reputation, geographic location, and network of connections. Therefore, some companies may find it easier to hire than others, which may skew the results from our snapshot of the quantum industry. Nevertheless, the companies interviewed reported that there were issues across three main types of job. 

Some companies reported it hard to find quantum information theorists who have a good understanding of the application of algorithms (33\% of all companies interviewed), though others found this less of a problem: ``a lot of the theory positions have been the hardest... it's very specialized as far as the number of groups in the world that focus on things like quantum computing algorithms or quantum error correction.'' This problem might be because theorists are often expected to ``hit the ground running'' being able to fill a specific role, and so fewer candidates are suitable. This expectation is likely a result of companies not having the resources (or economic need) to train quantum information theorists.

Hiring engineers with experience of analog electronics proved challenging (29\%): ``what's challenging is finding people with relevant electronics expertise, and by relevant I mean sort of a good blend of analog and digital.'' This is because there is less focus on analog electronics in electrical engineering degrees, so fewer people with these skills are entering the market, while there is still demand from other industries that rely on those same skills. Indeed, another interviewee, with a physics background, reported: ``you can't find any good analog electronic training anymore. The best are physicists on average.''

Finally, hiring senior level employees (19\%) with years of experience in the quantum industry and the ability to drive programs forward: ``it's hard to find somebody who has lots of experience and can significantly change the technical direction of the company.'' One reason for this is that ``it's a new industry and there's not very many people that have been around for more than a decade.''

\section{Discussion and Conclusions}\label{sec:discussion}

Higher education's role in preparing students for a career in the quantum industry has, so far, been dominated by physics departments. This is because it has mainly been in physics departments that proof-of-principle experiments have been carried out demonstrating quantum technologies. The quantum industry is interested in turning these experiments into products that can be sold to solve real-world problems. While theoretical contributions from mathematics and computer science departments are valuable, they only have a small contribution to the physical realization of a quantum sensor or quantum computer. This situation, of physics education dominating the curriculum vitae of employees in the quantum industry, is slowly evolving as the technologies mature into products that can be manufactured and sold. This means that there is an increasing need for engineers to refine these products to make them more reliable and lower cost. There is also the need to build a customer base for the hardware, such as people who know how to program quantum computers, so that companies, government, and academia can make use of the new tools that will become available in the future. The need for these latter users is entirely dependent on the timeline and success of the hardware efforts. There are, therefore, a number of places where there is an opportunity for higher-education institutions to play a new role in highlighting and providing the skills needed in the quantum industry to engineering and computer science students. Nevertheless, it is important to remember that the quantum industry still needs physics graduates at the bachelor's, master's, and Ph.D. level, who occupy a research and development role.

The results of our investigation suggest that classical skills in physics and engineering are valued just as much, if not more than, knowledge of quantum information science, for the majority of roles currently in the quantum industry. This is a reflection of the focus on the development of hardware, and the fact that quantum information theory lives in its own abstract space independent of the hardware. The difficulty is in how that theory is realized, such as managing to produce a $\pi/4$ phase shift on a qubit using a laser pulse. This requires classical skills, such as accurate control of the laser in terms of pulse duration, frequency, and intensity, as well as sufficient isolation of the qubit from its surroundings. Therefore, there is a lot of non-quantum-information-science knowledge that is needed: ``the science is ahead of the systems... we `know' how to build a fault tolerant quantum computer with a million qubits. Nobody knows how to make a hundred qubits.''

As the quantum industry is growing rapidly, there are a myriad of different job titles that describe the same or similar sets of responsibilities. One important example of this is the phrase ``quantum engineer(ing)'', which was mentioned by 29\% of the companies interviewed. In the discussion above, we deliberately avoided that phrase, as it has significantly different meanings to different people. For some, a quantum engineer is a Ph.D. physicist who has gone into industry, for others, it is a bachelor's engineer who now works in the quantum industry as a hardware engineer, and for others it is a software engineer who has studied quantum information science and now writes code for quantum computers. For each of these jobs, a different depth and breadth of skills are needed depending on the job and the company, in not only quantum physics but also discipline-based knowledge. Therefore, when a company says they need quantum engineers, it is important to clarify what is meant by that term.

There was a surprising lack of references to employees who have a computer science or a math background in the interviews. This is probably due to the current state of the quantum industry, but also the sampling of our study. If we assume our sample is representative, then the relative lack of computer science and math graduates in the industry is reflective of the hardware focus of the quantum industry, and also on the absence of training directed towards and awareness of job opportunities in the quantum industry in undergraduate courses. Then again, if students are aware of the opportunity, but are risk averse and recognize the nascent nature of the quantum industry, they may prefer to accept jobs working in classical computing, or other industries. Given that this study has been carried out by physicists, there is a possible bias in the phrasing of questions. We asked about employees who needed quantum knowledge, which led many interviewees to, initially, discuss only employees with a physics background. We discuss more about the limitations of the study in Section~\ref{sec:methods}. 

Similarly, machine learning has not appeared in our skills lists, despite 38\% of companies mentioning it, which is because it was not connected by the interviewees to any specific job or degree. Most of these companies described using machine learning to help analyze their data and optimize the design of their hardware. Only 14\% of companies mentioned quantum machine learning and not in any detail. As quantum machine learning is a use case for quantum computers, it should not be surprising that most companies do not look for employees with skills in it until the required hardware exists.

It is tempting to think that the natural home to study the application of quantum algorithms is in computer science or software engineering departments. To some extent that may be true, given the close relation of quantum information science to classical information science, however that is currently not the case. The conceptual and mathematical understanding developed in physics courses provides a foundation on which knowledge applicable to the quantum industry can be built. Hence, if computer science or software engineering departments want to include quantum computing in their curricula, there is still a large number of challenges in developing effective pedagogy to train students who are familiar with classical algorithms in the different paradigm associated with quantum algorithms. Examples of these differences are that quantum algorithms run in multi-dimensional vector spaces and give probabilistic outputs, as well as more practical considerations such as the inability to debug code by stepping through it while it is running. These differences were often highlighted in our interviews: ``there's the very basics of programming, but it changes so much when it comes to quantum programming. And then I think starting from scratch is almost as easy as retraining.''

Before concluding, we discuss an apparent discrepancy between the results we present from the desires of the quantum industry and the preparations being carried out by many higher-education institutions. We refer to the fact that the majority of higher-education institutions that are considering how they can adapt their existing programs to meet the needs set out in the NQI Act are considering introducing a master's degree (professional and/or traditional) in quantum information science~\cite{Kavliworkshop}, while only 14\% of the companies we interviewed mentioned the introduction of such degrees as a way for higher education to increase the workforce pipeline. A number of companies even suggested that all they would like is for classical engineers to have had a one or two semester quantum course, which would be significantly shorter and less expensive than a two-year master's degree. Given that the rate of supply of employees with Ph.D.s in quantum physics is not increasing, if companies wish to expand their workforce, most of the job growth will be in positions that require less quantum expertise. While a master's in quantum information science might help an applicant stand out during the hiring process, we recall that for these types of positions, companies reported that they valued expertise in engineering skills over depth in quantum knowledge. It is essential, therefore, that there is communication between higher education and the quantum industry over the coming years in order that each can adapt and prepare for the changes that are taking place in the other.

The qualitative nature of this work provides clarification on what the quantum industry is, the types of jobs within it, and what skills and knowledge are currently valued by the industry. However, there remains a number of unanswered questions, which are outside the scope of the present work. One key question is quantifying the number and size of companies within the quantum industry, as well as the distribution of jobs within the industry. While that has not been the goal of this work, the qualitative descriptions of the quantum industry and the jobs within it provide a basis for the construction of a vocabulary and definitions that can be applied in future research and policy work. This is important, as there is currently no agreed upon terminology, as we have discussed, for the types of jobs within the quantum industry.

Finally, there is the question of what changes can a higher-education institution make that would benefit their students? There is not one correct answer here, however one thing that a higher-education institution could do is introduce an intro-level quantum course focusing either on the hardware or algorithms aspects of quantum information science. Such a course would have appeal as both general interest as well as be useful for a variety of science, technology, engineering, and math majors (see the similar recommendation by the Defense Science Board\cite{DSB2019}). When developing a new course, or even a larger program, the breadth of the quantum industry means that choices must be made: what area of the quantum industry should it focus on, sensors, networking and communications, or computing? Should it be a hardware focused course, with hands-on activities? Or more abstract, focusing of quantum programming, or pure quantum information theory? Who are these courses for, students or professionals? In which department should these courses be given? These choices should be based on the expertise available at that institution, the needs of the students, and consideration of the local and national connections to industry of the institution. The latter dependency can be important, as it not only provides routes for students into industry, but can also provide resources for the university in developing these new courses, such as access to quantum computers, training content, as well as possible directions for collaboration on training hardware that could be hosted at the higher-education institution. We hope that our results and suggestions for course content in Table~\ref{fig:Real-worldQIS} and the Supplementary Tables~\ref{tab:trad-q-theory}-\ref{tab:optics} can help provide some guidance when faculty are considering these questions.


\begin{acknowledgments}
We appreciate the assistance of the QED-C with the recruitment process for this study and thank all interviewees for contributing to this work. We thank Jessica Hoehn and Mike Verostek for carrying our inter-rater reliability on the coding of interview transcripts and Mike Verostek for helping with the categorization of skills. We thank Noah Finkelstein and David Steuerman for their comments on the manuscript. \textbf{Funding:} This work was done as part of the CUbit Quantum Initiative, which includes Q-SEnSE: Quantum Systems through Entangled Science and Engineering (OMA-2016244). M.F.J.F. and H.J.L. were supported by the University of Colorado Boulder and by NSF PHY-1734006. B.M.Z was supported by Rochester Institute of Technology and by NSF MPS-1937076. \textbf{Author contributions:} All three authors designed the research. M.F.J.F conducted the interviews, performed the coding analysis, and resulting statistical analysis. All three authors discussed results and wrote the paper. \textbf{Competing interests:} The authors declare that they have no competing interests. \textbf{Data and materials availability:} All data needed to evaluate the conclusions of the paper are present in the paper. Additional data related to this paper may be requested from the authors. 
\end{acknowledgments}

\newpage

\setcounter{section}{0}

\renewcommand\thesection{S.\Roman{section}}
\renewcommand\thesubsection{S.\Alph{subsection}}
\renewcommand\thesubsubsection{S.\arabic{subsubsection}}
\setcounter{table}{0}
\renewcommand\thetable{S.\Roman{table}}

\pagenumbering{roman}

\clearpage

\section{Skills needed by job. }\label{sup:skills-by-job}
In this section, we extend the discussion of Section~\ref{sec:career-opportunities} by providing descriptions of each job role and how the skills required by them contribute to the quantum industry. The percentage given after each skill below is the percentage of the occurrence of that skill from the number of companies (N) that stated they had employees of the given job type. We do not report percentages that are low enough to be able to identify individual study participants.

\subsection{Engineer (N = 20). }
Engineers generally have a bachelor or master's degree and usually have earned either a traditional engineering degree or a physics degree. Below, we detail the five main types of engineer. Generally, an engineer's expertise in designing efficient systems that are also easy to manufacture is prized above any knowledge of quantum physics or quantum information science. Engineering roles typically take the specifications given by an experimental scientist and use their engineering domain knowledge and experience to design and deliver the required component. 

\subsubsection{Electrical engineer (N = 16). } The majority of the engineering jobs are electrical engineers. Electrical engineers form an important part of hardware teams and there is demand for different specialist knowledge depending on the specific hardware, such as the design and manufacture of microwave circuits that are used to manipulate qubits in superconducting hardware or radio frequency circuits to probe atomic transitions (38\%). Electrical engineers play a role in constructing precision control and feedback systems (31\%), so that measurements of delicate quantum states can be made, which requires an understanding of analog electronics (31\%). Another example is the need for specialists in building low-noise power supplies for laser systems.

\subsubsection{Software engineer (N = 11). } Software engineers in most quantum companies are involved in developing the classical code that runs a quantum device: from control systems (27\%) for the hardware of quantum sensors (such as reading data using FPGA technology), to interfaces and programming languages for end users of quantum computers (27\%). Indeed, when considering the skills needed for the control of hardware, there is some significant overlap with electrical engineering, such that many companies do not necessarily make a distinction between the two. Additionally, on the hardware side, more knowledge of the practical system is required, and some of the skills required by a software engineer might also be expected of in an experimental scientist. Moving away from the hardware to the development of a programming language for a quantum computer, this requires a slightly different set of skills and might be completed by a software engineer with a computer science background. Coding for classical systems (82\%) was reported by most companies, by a large margin, as the skill needed for software engineers, and they are valued for their experience in this area.

\subsubsection{Mechanical engineer (N = 9). } A mechanical engineer is involved with the design and construction of hardware to enable quantum information processing. As, for example, the ability to process photons with low loss and noise levels is essential for the precise measurements required in a number of different systems in the quantum industry (e.g., probing atomic transitions and measuring entangled photonic systems), it is important to be able to manufacture support systems for optical apparatus (56\%) with low vibrations to reduce noise (22\%). Additionally, mechanical engineers may be employed to achieve the isolation required for high-coherence times, and hence require knowledge of how to build systems to operate in vacuum (22\%) and at cryogenic temperatures (22\%). In all of these tasks, experience of mechanical design (44\%), such as geometric dimensioning and tolerancing, is valued by many hardware companies. More experienced, or master's degree level mechanical engineers, may be tasked ``to do more system design'', such as how the different components of a system fit together. This task would often be shared with the experimental scientists, who would have written the specifications for the device.

\subsubsection{Optical engineer (N = 6). } Optical engineers can work closely with mechanical engineers on systems as described above (as opto-mechanical engineers) (100\%). Additionally, optical engineers would be expected to have experience using commercial software packages to design and model beam propagation (33\%). Understanding and using laser systems (50\%), such as nonlinear feedback in a laser cavity, locking to a cavity, and the production of narrow laser linewidths, are valued skills, especially with the often unique specifications (wavelengths) required to probe specific atomic transitions (33\%). Experience with photonic integrated circuits is also valued by some companies, which is a skill set that overlaps with electrical engineering.

\subsubsection{Systems engineer (N = 5). } Systems engineers are expected to have not just ``a really good understanding of the quantum aspects'' but ``as many of the aspects as possible'' of the entire device. Due to these requirements, a background in physics is normally good preparation for this role, rather than a pure engineering background. This job is slightly different from an experimental scientist described below, as it requires engineering project management skills, rather than a more research-based approach to developing a project.

\subsection{Experimental Scientist (N = 18). }
Experimental scientists often occupy job roles as ``Quantum Physicists'' or ``Quantum Engineers'' and typically have a Ph.D. in a relevant sub-field of physics (atomic, molecular and optical physics) for ion/atom trapping technologies; superconducting hardware; or photonics). They are normally working with a team of other Ph.D. experimental scientists, as well as employees with an engineering background to develop a new piece of hardware, often taking on a leadership role within that team. The number one skill required of experimental scientists is that of laboratory skills and experience (67\%), composed of: generic electronics skills (56\%); statistics and data analysis (50\%); coding of classical systems - mainly to control experiments or process data (44\%); modeling (39\%); and troubleshooting (33\%). The expectation of quantum knowledge is limited to specific knowledge on qubit hardware (39\%) that is being used, as well as the Hamiltonian description of the system (33\%). Other skills required by experimental scientists, but not included in the above list because of the hardware dependent nature of those skills are: optics and opto-mechanics (56\%); lasers (50\%); knowledge of atomic energy levels and atom-photon interactions (39\%); and knowledge of radio frequency or microwave systems (33\%). We separate these latter skills to emphasize that an experimental scientist is expected to have a breadth of skills related to experimental science and working in a lab, but also a depth of skills and knowledge in a specific area relevant to their job. These latter skills reflect more on the convolution between the sampling of companies from the quantum industry in this study with the types of activities being undertaken within the quantum industry, and we do not attempt to deconvolve these two.

\subsection{Theorist (N = 13). }

There are two different subsets of theorists that work in the quantum industry. The most prevalent is the quantum information theorist, though a distinct group of theorists specializing in modeling hardware and the practical realization of quantum information theory also exists.

\subsubsection{Quantum information theorist (N = 13). }
Quantum information theorists work mainly in quantum computing companies (both hardware and algorithms and applications) and tend to have a Ph.D. They are required to have a good understanding of quantum algorithms (85\%). Some of them are developing new algorithms and hence need an understanding of what a qubit is in the abstract (15\%) as well as linear algebra (23\%). This latter percentage is surprisingly small, given that linear algebra is a prerequisite for studying quantum physics. A likely reason is that knowledge of qubits and linear algebra for a theorist was implicitly assumed by the interviewee and so was not mentioned in the interview and could not be picked up in our analysis. Other quantum information theorists are figuring out how to implement the algorithms with hardware, therefore, they also need an understanding of the gate model of quantum computing and quantum circuit design (38\%), as well as the Hamiltonians that describe the system (31\%). A possible specialization for quantum information theorists is that of error correction (38\%), which is essential because of the noise associated with physical implementations of quantum computers. Finally, knowledge and experience with coding quantum systems (38\%) generally is an important skill, which relies on coding classical computers to interface with and run operations on a quantum computer. Though these skills overlap somewhat with the application researcher job, the distinction between these two jobs is whether the employee is engaged or not with external clients.

\subsubsection{Theorist - not quantum information science (N = 4). }
This category of theorist, while still having a background in abstract quantum information science, focuses on modeling the hardware (50\%), such as qubits, sensor atoms, optics, etc. that is being developed by a company and considers how to optimize that system. They usually have a Ph.D. As such, they generally have experience and knowledge of material properties of the system (50\%) as well as practice with describing a system through its Hamiltonian (50\%). This job is somewhat specialized, contributing to its low occurrence in our study, as these modeling skills are also aspects of an experimental scientist's role.

\subsection{Technician (N = 9). }
Technicians work in the quantum industry to manufacture devices or operate systems through well-defined processes. They usually have a high-school diploma or an associate degree. Examples may include production of integrated circuit boards (22\%), mounting and alignment of optical systems (33\%), cooling of cryogenic systems, or baking of vacuum systems. As these skills are already valued outside of the quantum industry, respondents did not focus on the skills required of technicians, beyond looking ``for the disposition... we'd like to find people who are careful and patient and have training doing delicate [things].'' Indeed, technicians do not need to ``have a lot of savvy when it comes to quantum mechanics'', as they are given a well-defined task and have to complete that task to a certain standard.

\subsection{Application researcher (N = 7). }
An application researcher's role is to work with clients who want to learn whether their problems can be solved using a quantum computer. Application researchers range from bachelor degree holders to having postdoctoral experience. In this role, employees need a good understanding of quantum algorithms (86\%), with the minimum knowledge being: what algorithms exist, what they can do, and how they can be used. Employees do not necessarily need the mathematical skills to develop new algorithms, however, this is a company dependent requirement, and there is still a strong desire for employees to be well versed in linear algebra (57\%). These employees are also expected to have knowledge of the different hardware approaches to quantum computing (57\%), and as such which may be best to run possible algorithms on. Furthermore, these employees might be expected to write and run demonstration code on quantum computers (57\%). From our sample, we found an equal number of companies with employees using quantum annealing (43\%) and gate-model quantum computing (43\%) systems. In all cases, a good knowledge of statistics, data analysis, and noise sources (29\%) is needed for troubleshooting and debugging quantum code (29\%). Another aspect that is desired for application researchers is some knowledge or experience in a second field, such as chemistry, that may benefit from the application of quantum computing.

\section{Skills valued from degree subjects and levels. }\label{sup:skills-by-degree}
In this section, we describe how each degree (subject and level) provides skills that are valued by the quantum industry, extending the overview provided in Section~\ref{sec:employees-gained-skills}.

\subsection{Physics. }
\subsubsection{Ph.D. } {\em``a large number of our people are physics Ph.D.s''}

The largest contribution to the quantum industry, of all degree subjects and levels, is from employees with a Ph.D. in physics (Fig.~\ref{fig:OnlySubjectLevel}). These employees mainly contribute to the experimental scientist, theorist, and application researcher jobs. The reasons why a Ph.D. in physics is often a requirement for these jobs are:
\begin{enumerate}
    \item Experience in delivering on a scientific project.\\
{\em ``generally, what they bring is... the ability to drive a research program.''}\\
This encompasses project design, troubleshooting, and how to prove that a device does what is claimed of it. While these skills can be gained in other disciplines, a background in physics provides a shared language that companies value.
    \item Deep knowledge of a specific quantum system and latest results.\\
{\em ``somebody who can read scientific papers, somebody who's comfortable interacting with professors of research groups and... [using] very specialized language''}\\
{\em ``the Ph.D. is basically like hiring somebody who's already been doing the work for six or eight years''}\\
To be more specific, we consider experimental and theoretical Ph.D. programs separately. For experimental Ph.D.s, this specific quantum knowledge is how `quantum' manifests itself in the real world - how to actually perform measurements of a quantum system to extract information out of it. For theoretical Ph.D.s, this refers to being comfortable with the mathematics behind quantum information theory and/or the theory that describes the physics of a real quantum system (i.e., including noise). Having completed a Ph.D. in a specialist sub-field of quantum physics, it is relatively easy for someone to transfer those skills to similar work in a company that is commercializing that technology, which are typically the experimental scientist and theorist jobs. Regarding the application researcher, these roles are more relevant to someone who has completed a Ph.D. in quantum information theory and algorithm development, who desires a career more grounded in the application of that theory to solve real-world problems. Nevertheless, for application researchers, having completed a Ph.D. in quantum information theory is not a necessary requirement for this job (see Fig.~\ref{fig:DegreeSubjectLevel}). 
\end{enumerate}

The relatively large number of engineering jobs that appear in Fig.~\ref{fig:DegreeSubjectLevel} for Ph.D. physicists can be attributed to the larger role engineering takes in industry compared to academia, as the ultimate goal is to produce a functioning product that can be sold. Additionally, there is no strict convention that distinguishes an engineering from a physics role, therefore allowing considerable variation in the use of these terms between companies (see Section~\ref{sec:discussion}).

\subsubsection{Master's. }
Given that terminal master's degrees in physics are uncommon (at least for the time being - see the following section on what training and education programs would be helpful), these master's degree holding employees are more likely to have started a Ph.D. program, but decided to leave with a master's and enter industry. Therefore, the extent to which an employee with a master's would be suitable to fill a research type job depends on the extent of the research experience they received during their degree. We see in Fig.~\ref{fig:DegreeSubjectLevel} that the distribution of jobs for physics master's degree holders is more similar to bachelor's degree holders than employees with Ph.D.s.

\subsubsection{Bachelor's. }
Neither of the two key skills discussed for Ph.D. physicists are often evident in undergraduate degree holders, as they are developed through conducting a research project over an extended period of time. The basic knowledge of quantum mechanics developed in an undergraduate course provides a familiarity to the language used in the quantum industry, but is limited in its usefulness: ``if somebody's taking the quantum mechanics class then they haven't really had to think about how to use quantum mechanics in the context of doing a project.'' Therefore, where bachelor-level physics-degree holders are employed in the quantum industry, they often have engineering roles (Fig.~\ref{fig:DegreeSubjectLevel}).

One limitation of the data presented in Fig.~\ref{fig:DegreeSubjectLevel} is that it does not account for the experience of an employee. For example, they may have an undergraduate degree in physics, but have been in the company for a number of years and gained the necessary experience to move into an experimental scientist or senior engineering job. As such, ``the biggest difference [between a bachelor's and a Ph.D.] would be independence and so that would really be both knowledge and thought process'', which can also be gained through experience in employment.

\subsection{Engineering. }
\subsubsection{Ph.D. } ``when we've interviewed Ph.D. engineers... it's actually sort of a negative at times because a Ph.D. is so myopic... the ones that for us have been most successful are the ones that then took that skill to industry and applied it to customer specific system.'' The relative lack of employees with a Ph.D. in engineering compared to in physics reflects the tendency for engineers to enter industry after completing a bachelor's or master's degree and not to continue in academia~\cite{Roy2018,Mulvey2017}. While engineering Ph.D. employees would most likely share with Ph.D. physicists experience in delivering on a scientific a project, they may not have the deep knowledge of a specific quantum system. The data in Fig.~\ref{fig:OnlySubjectLevel} and Fig.~\ref{fig:DegreeSubjectLevel}, suggest a relatively large number of Ph.D. engineers, however, many interviewees referred to ``physics and engineering Ph.D.'' jobs which could be interpreted as either a job requiring a physics or engineering Ph.D. or a job that crosses the border between physics and engineering. In the systematic way our coding analysis has been applied we are unable to disentangle this subtlety in the numerical analysis. Based on the broader context of the interviews, where the majority of discussions were about employees with Ph.D.s in physics, we surmise that the counted number of Ph.D. engineers in these sets of data is an overestimate (see the discussion of limitations in Section~\ref{sec:methods}).

\subsubsection{Master's and bachelor's. }
We group together both master's and bachelor's degrees in engineering, as the majority of jobs occupied by employees with these qualifications are engineering roles. The difference between the bachelor's and the master's degrees are a matter of specialization that occurs during the master's course, for example: ``working in a lab or a university where they have [a course] to do with refrigerators.'' Nevertheless, the requirements of an engineering job in the quantum industry appear to be well satisfied by a traditional undergraduate degree in electrical, software, mechanical, or optical engineering. Companies interviewed value engineers for their expertise in manufacturing and production of components: ``I don't expect a Ph.D. physicist to be a good person to do a design for an ultra-low noise current source. I'd prefer that work to be done by an electrical engineer who's got a stronger background, because it ends up being cheaper and faster to do it that way.''

\subsection{Computer science. }
\subsubsection{Ph.D. }
A Ph.D. in computer science lends itself to job opportunities in both theorist and application researcher roles. This is, similar to a Ph.D. in physics, clearly dependent on the topic of the Ph.D. dissertation. These roles are in quantum computing, where the quantum logic is abstracted from the hardware, and hence little knowledge of the physics is required. Another specialty that a computer science degree provides to the quantum industry is in high-performance computing, which is used to simulate quantum systems for understanding sensors and networks, as well as to test whether a quantum computer is working as expected. The fact that few companies in our sample reported hiring computer science Ph.D. holders could indicate either a lack of demand for quantum algorithm theorists by the quantum industry; a lack of supply from computer science departments; a bias in the hiring practices of the companies in our sample; or a bias in our sampling of the quantum industry (see Section~\ref{sec:methods}). At this moment, our data is not sufficient to determine which of these options may be the case, so we can only postulate that the hardware focus of many companies interviewed limits the need to hire a large number of quantum algorithm experts.

\subsubsection{Master's and bachelor's. }
At both the master's and bachelor's degree levels, the majority of jobs are as software engineers, to help run the back-end of the quantum system, for example, designing controls for hardware, or compilers that translate high-level code into operations on a quantum computer. There are, however, some computer scientists at the bachelor's degree level employed in theory or application researcher roles, which is based on their experience with computational logic, as well as having a passion for quantum computing: ``there's... people within the group that have computer science backgrounds that are kind of self-taught physicists''.

\subsection{Chemistry. }
\subsubsection{Ph.D. }
Advanced chemistry degrees provide domain knowledge that may be used in one potential use case for quantum computers: the simulation of molecules and reactions. Additionally, the field of physical chemistry has a number of overlapping technical areas with relevant physics for quantum sensing and computation, such as methods for trapping and studying single atoms (e.g., ion traps, laser cooling, and magneto-optical traps).

\subsection{Math. }
\subsubsection{Ph.D. }
Similarly to having a Ph.D. in computer science or theoretical physics (particularly quantum information science), a Ph.D. in mathematics with a focus on the math of quantum information theory provides excellent background knowledge for work in algorithm development for a quantum computing company.

\section{Training provided by companies by job type}\label{sup:training-provided-by-companies}
In this section, we elaborate on the discussion in Section~\ref{sec:training-provided-by-companies} by breaking down the training companies provide on-the-job to their employees based on the type of job they have.

\subsection{Experimental scientists} Experimental scientists are trained on the domain knowledge associated with the specific hardware that is being developed by the company. This is knowledge that can be learned only on the job because the details of the configuration of equipment will be known only by those already working on it. Having completed a Ph.D. in an area of physics well matched with the activities of a company, a new employee is well suited to the style of working in the quantum industry, as many of the employees have similar backgrounds, with a number of companies having been spun out of university research groups. The expectation that employees will undertake independent learning to get up to speed, therefore, would be a familiar expectation to new employees. However, there are two areas where companies have reported the need for additional training: in the principles of engineering and system design; and in standards for software development and writing code collaboratively. There already exist many providers of software development courses, be they through online courses or day-to-week long workshops, which companies bring in for their employees. Examples of topics employers expect employees to know are ``how to collaboratively write code with code review... write unit tests... know data structures and how useful API might look like.'' For engineering principles, most companies reported that employees would be expected to learn through interacting with employees with engineering backgrounds: ``We need those kind of people with that professional engineering skill, that's something that these atomic physicists can learn.'' It was highlighted that familiarity with writing specifications that could be sent to a machine shop to be manufactured and that made sense to an engineer or technician was beneficial; one interviewee with a background in physics stated: ``where physicists tend to suffer is that they're always really bad at writing specifications.''

\subsection{Theorists} Theorists are generally in a similar situation to experimental scientists, in the sense that the work they would be doing in the quantum industry is well matched to their academic (Ph.D.) background. However, because the nature of their work does not depend so strongly on proprietary knowledge of hardware as the work of an experimental scientist, there is a greater expectation that they ``can hit the ground running because [they've] just done a Ph.D. and... what [they] work on will be pretty closely related to what [they] have been doing previously.''

\subsection{Engineers}
Engineers are valued for their expertise in their specific sub-discipline of engineering. The on-the-job training for engineers generally has three aspects: one is becoming familiar with the product and processes of the company, be it learning to use the specific software packages a company uses to design, for example, optical systems, or getting hands-on experience with the hardware; the second is gaining some understanding of the physics of the system; the third is developing manufacturing related skills. The first of these is similar for all employees, which we have already discussed.

The second aspect refers to the ability of employees to function effectively in the quantum industry. As such, it helps if engineers are familiar with the language and specifications associated with quantum technologies: ``so we take classical experts... and we get them to be familiar with quantum constraints and requirements'' to become ``pretty high functioning, have a good literacy, even if they wouldn't be able to talk in very mathematical language about it.'' Different companies approach the teaching of this quantum awareness in different ways. Usually, these are short courses provided in-house by senior members of the company. For example, one company implemented ``some formal sort of meetings/classes where we give some background, especially to the non physicist''; while another brought in an external collaborator to deliver ``roughly 20 lectures'' specifically to engineers where ``they're going to come out of it with enough of the basic understanding to then turn around and be doing the kind of work I'm looking for''.

The third aspect is a form of continuing professional development. Manufacturing skills are highly valued, and making sure engineers are fully aware of best practices can save time and money for a company, so training in ``geometric dimensioning and tolerancing'' through a community college class, is useful for hardware companies. An alternative method of training reported was ``bringing coaches... into the company and we've structured a three month training program... two times a week''.

Finally, another possibility for training engineers is to go back to school in order to get a higher degree, for example, one engineer, after spending years working in the quantum industry, ``wanted to go get his own Ph.D. in physics''. This can also be the case for junior experimental scientist roles, where the employee might have entered the company from a bachelor's degree. Depending on the resources available to a company, they may support an employee in gaining further academic qualifications, through part-time work arrangements or direct financial support. However, these plans require the costs and benefits associated with each individual situation to be assessed by the company.

\subsection{Technicians}
The training provided to technicians is in the ``assembly and test procedures'' that they would be carrying out, and requires a disposition and a specific set of skills, as discussed earlier when describing this job. In terms of training on knowledge about how the system works, that would be more of a curiosity for technician roles rather than essential to the job: ``people working on the fabrication side of things where they're looking to just build that, but they don't necessarily need to understand [quantum].''

\subsection{Application researchers}
Application researchers are interesting regarding their training needs, as the academic backgrounds of these employees vary greatly. However, because of the abstract nature of quantum computation, provided that an employee has a sufficient background in linear algebra and the motivation to learn, there is not a significant amount of physics knowledge required to understand the computational process. The main training content that employers provide for these employees is in the practicalities of programming a quantum computer using one (or more) of the many available languages. This training comes in many different forms, including online courses provided by a university or a company: ``a couple of our programmers did take... online certifications''; or ``three or four days worth of training courses with our experienced professionals... as an introduction to quantum computing and the principles behind it''; or even just through independent learning from textbooks: ``We gave them a couple of books.'' In addition to short courses on programming quantum computers, these employees are often supported through ``one-on-one mentoring'' to be able take the basics learned from a short course and apply it to a problem of relevance to the company. 

Separate from training in the pure aspects of quantum computation, there is also a need for employees to get an understanding of the systems on which their codes run. This is because current hardware is not perfect, so some knowledge of the system is needed to understand how algorithms may fail. For this, training where employees ``get walked through the entire stack [from the hardware, the back-end software, to front-end user interface]'' is often provided.

\clearpage
\onecolumngrid
\section{Interview protocol}\label{sup:interviewprotocol}

\subsection*{Part 1 - Company context}
\begin{enumerate}
    \item How does the company fit within the Quantum Industry?
    \item What products does the company currently produce, if any?
    \item How many people work on quantum-related technologies at the company?
    \item What is your role within the company?
    \item How many people do you have responsibility for (i.e. in your team)?
    \item Does your company currently have structural ties to higher education? Could you describe those connections? What do they entail? (maybe on an advisory board? Researchers? Adjunct professors, etc.?)
\end{enumerate}

\subsection*{Part 2a - Skills as a function of academic preparation}
In the following questions I want to ask about skills and knowledge related to quantum-specific job opportunities at your company. 
\begin{enumerate}
    \item Firstly, what kinds of job opportunities are (or have been) available at your company? (Need to agree on specific jobs that will be talking about before proceeding.)
    \item Considering those job opportunities you just mentioned, what level of formal education do you expect employees to have for each position? Do you hire Associates-level, BS, Masters, PhD?
    \item We realize there may be substantial differences in prerequisite skills depending on the position. 
    \begin{enumerate}
        \item What are the differences between a BS and a Masters/PhD position in terms of job description?
        \item What are the differences in terms of key skills and knowledge expected coming in?
    \end{enumerate}
    \item (if relevant) Are BS-level positions ever promotable to higher skill jobs?
    \begin{enumerate}
        \item If so, does that happen through internal training or do employees go back to school?
    \end{enumerate}
\end{enumerate}

\subsection*{Part 2b - Specific skills and knowledge}
\textit{Remembering to break this down by kinds of jobs available at the company.}

\textbf{Quantum-specific knowledge}
First, I want to ask about quantum-specific knowledge that employees in your company need to know, then I will ask about scientific and technical skills required in your company. 
\begin{enumerate}
    \item What quantum-specific knowledge do employees at your company need to know; for example linear algebra, Hamiltonians, the Schroedinger equation, information theory, probability theory?
    \begin{enumerate}
        \item How often would an employee use such knowledge (daily/weekly or only as needed/rarely)?
    \end{enumerate}
    \item 	What physics knowledge do employees at your company need to know; for example conservation laws, symmetry, models and their limitations?
    \begin{enumerate}
        \item How often would an employee use such knowledge (daily/weekly or only as needed/rarely)?
    \end{enumerate}
    \item What knowledge of practical systems do employees at your company need to know; for example properties of optical systems, materials, electronics?
    \begin{enumerate}
        \item How often would an employee use such knowledge (daily/weekly or only as needed/rarely)?
    \end{enumerate}
    \item What quantum-specific knowledge do you expect employees to have prior to employment? What knowledge do you expect employees to gain during employment?
    \item Do you notice any trends in the strengths or weaknesses of quantum-specific knowledge of new hires? What are the trends?
    \item What training opportunities, if any, exist to help employees develop the required knowledge on the job? Is that formalized in any way?
\end{enumerate}

\textbf{Scientific and technical skills}
\textit{Remembering to break this down by kinds of jobs available at the company.}
Now I want to talk about scientific and technical skills, for example solving ill-posed problems, using instrumentation, troubleshooting apparatus, competency in using off-the-shelf software, coding, performing statistical and uncertainty analysis.
\begin{enumerate}
    \item What scientific and technical skills do employees at your company need to know? (Are there any additional skills?)
    \item How often would an employee use such skills (daily/weekly or only as needed/rarely)?
    \item Which of these skills do you expect employees to have prior to employment? Which do you expect employees to gain during employment?
    \item Do you notice any trends in the strengths or weaknesses in scientific or technical skills of new hires? What are the trends?
    \item What training opportunities, if any, exist to help employees develop these skills on the job?
\end{enumerate}

\textbf{Tidying up.}
\begin{enumerate}
    \item Are there any of these skills or knowledge that are unique to your company?
    \item Are there any ``soft'' skills (leadership, communication, multitasking, etc.) you think are especially relevant? 
    \item Are there any other quantum-specific knowledge or scientific and technical skills that you think are required to work at your company that we have not discussed?
\end{enumerate}

\subsection*{Part 3 - Training and hiring}
\begin{enumerate}
    \item When you advertise for a position do you see a lot of unqualified applicants?
    \begin{enumerate}
        \item If yes, what fraction? How many in total? 
        \item What are the most common deficiencies you see in applicants?
        \item What are the hardest to fill positions? Why?
        \item Would you say that there is a gap between the skills and knowledge employers are looking for and the skills and knowledge of new graduates?
    \end{enumerate}
    \item (if not answered already) Thinking of recent hires, what tend to be the areas where new hires need training?
    \item What skills are hardest to train? Why?
    \item What would be the biggest improvement that a university could make to help their students succeed in the new job?
    \item If there are gaps in training or knowledge of an employee, how is that viewed and handled within the company? How is training then provided (1-1, online, workshops, external courses, mentoring)?
    \item What would the best way that a university could provide training to fill those gaps (Online learning (e.g., MOOCs)? Short courses? (Remote) Masters?)?
    \begin{enumerate}
        \item What is the maximum amount of time you would expect an employee to spend on such training?
    \end{enumerate}
    \item (Going back to hiring) Where do you post job descriptions? How do you find people who may be good candidates?
    \item What parts of the application do you find most useful in early screenings (e.g. cover letter, transcripts, letters of recommendation)?
    \item What criteria do you use to identify the most promising applicants from their written application materials?
    \item What kinds of tasks/questions do you ask applicants at interview to decide who should be moved forward in the application process (for example solving a math problem, or coding something)?
\end{enumerate}

\subsection*{Part 4 - Wrap-up}
\begin{enumerate}
    \item Great, thanks! Those were all the specific questions I had. 
    \item Is there anything else relevant to education and workforce development related to the Quantum Industry that you would like to tell me?
    \item Is there anybody else or any specific companies that you would recommend that I should contact to invite to participate in this research study?
\end{enumerate}

\clearpage
\section{Courses and their relationship to skills valued by the quantum industry}\label{sec:skillstables}
\begin{table}[h]
    \centering
    \caption{Skills and examples of relevance to the quantum industry for a possible traditional quantum theory course.}
    \includegraphics[width=0.9\textwidth]{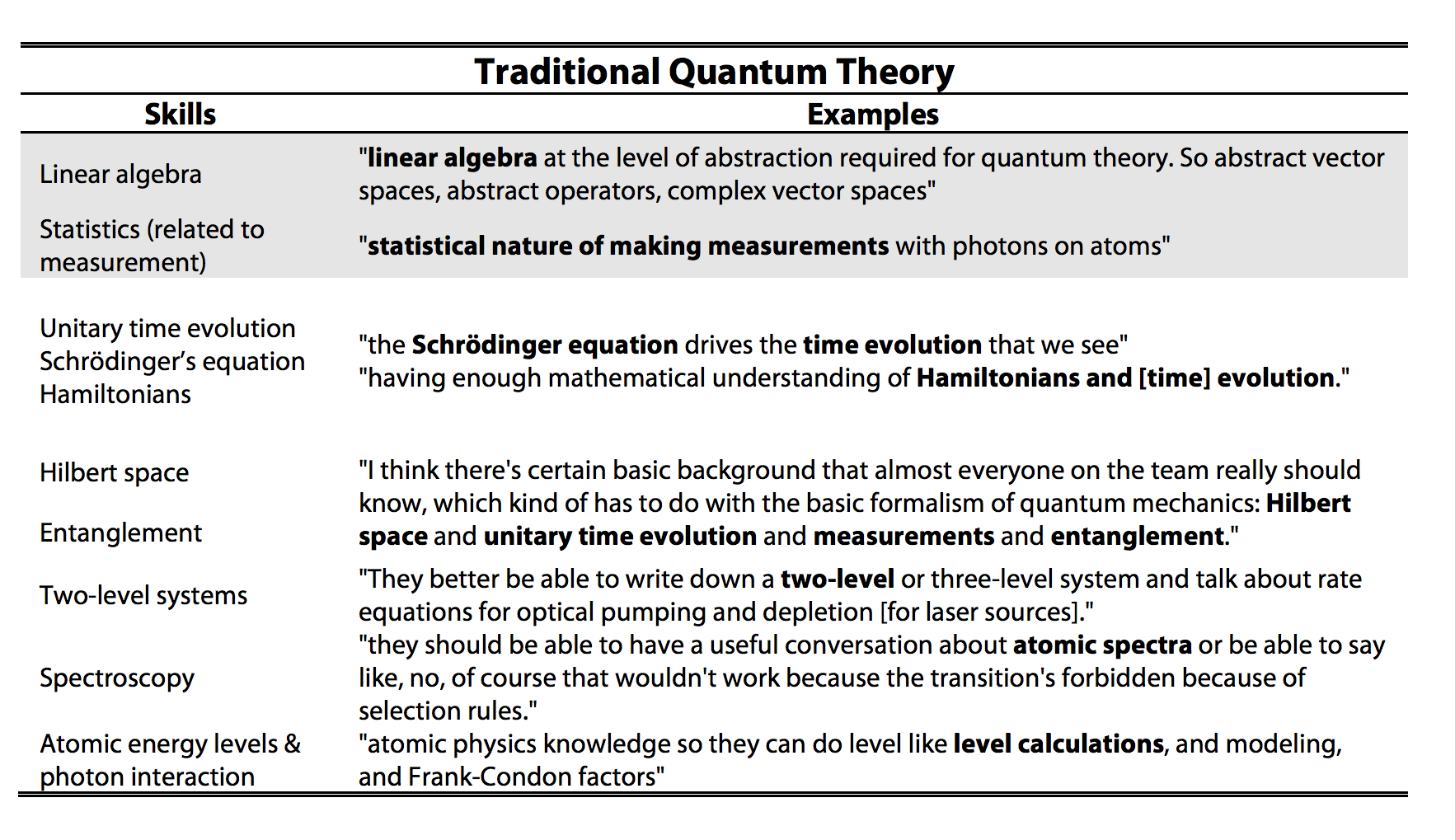}
    \label{tab:trad-q-theory}
\end{table}
\begin{table}[h]
    \centering
    \caption{Skills and examples of relevance to the quantum industry for a possible quantum information theory course.}
\includegraphics[width=0.9\textwidth]{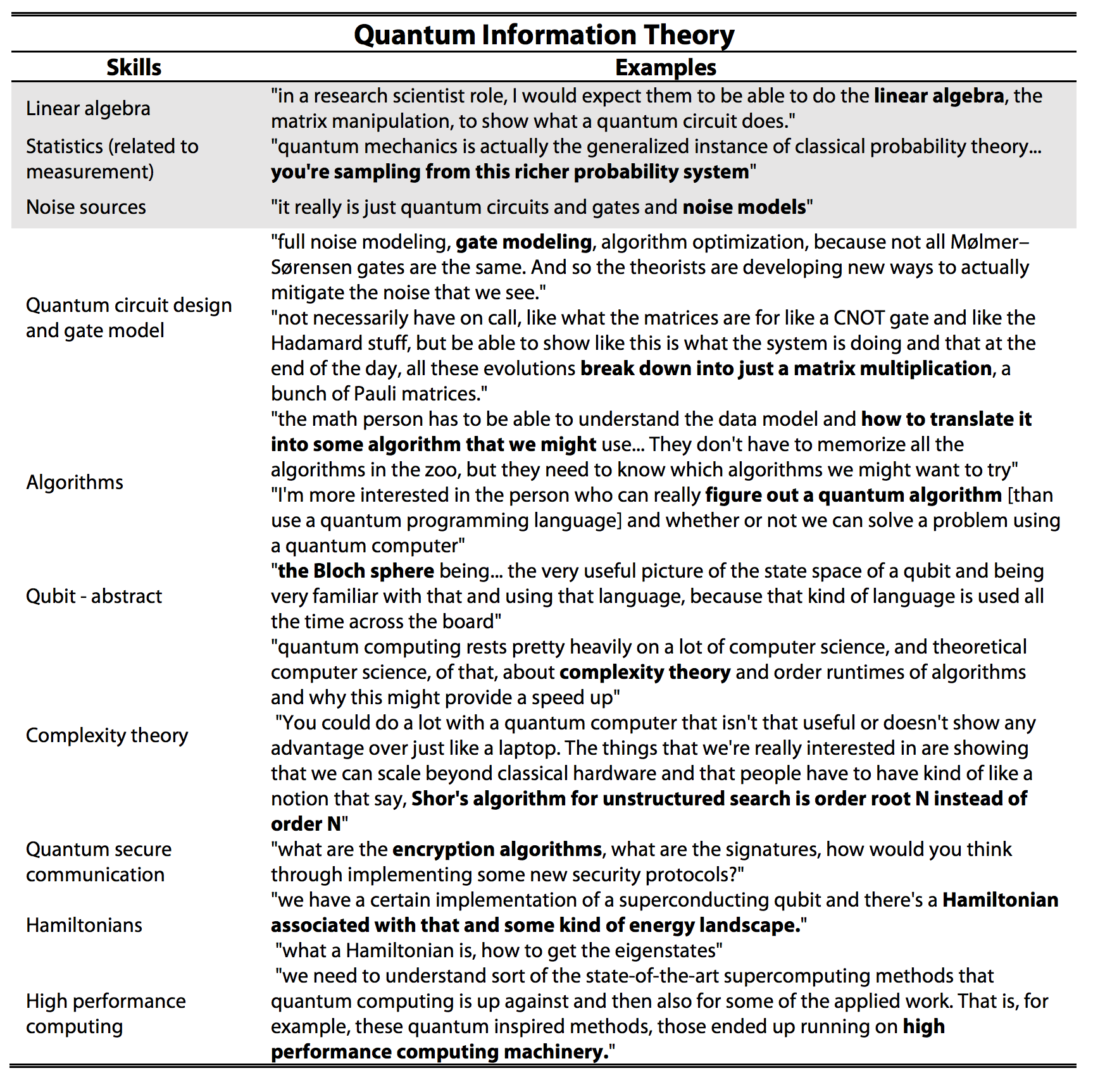}
    \label{tab:qis-theory}
\end{table}
\begin{table}[h]
    \centering
    \caption{Skills and examples of relevance to the quantum industry for a possible hardware for quantum information course.}
\includegraphics[width=0.9\textwidth]{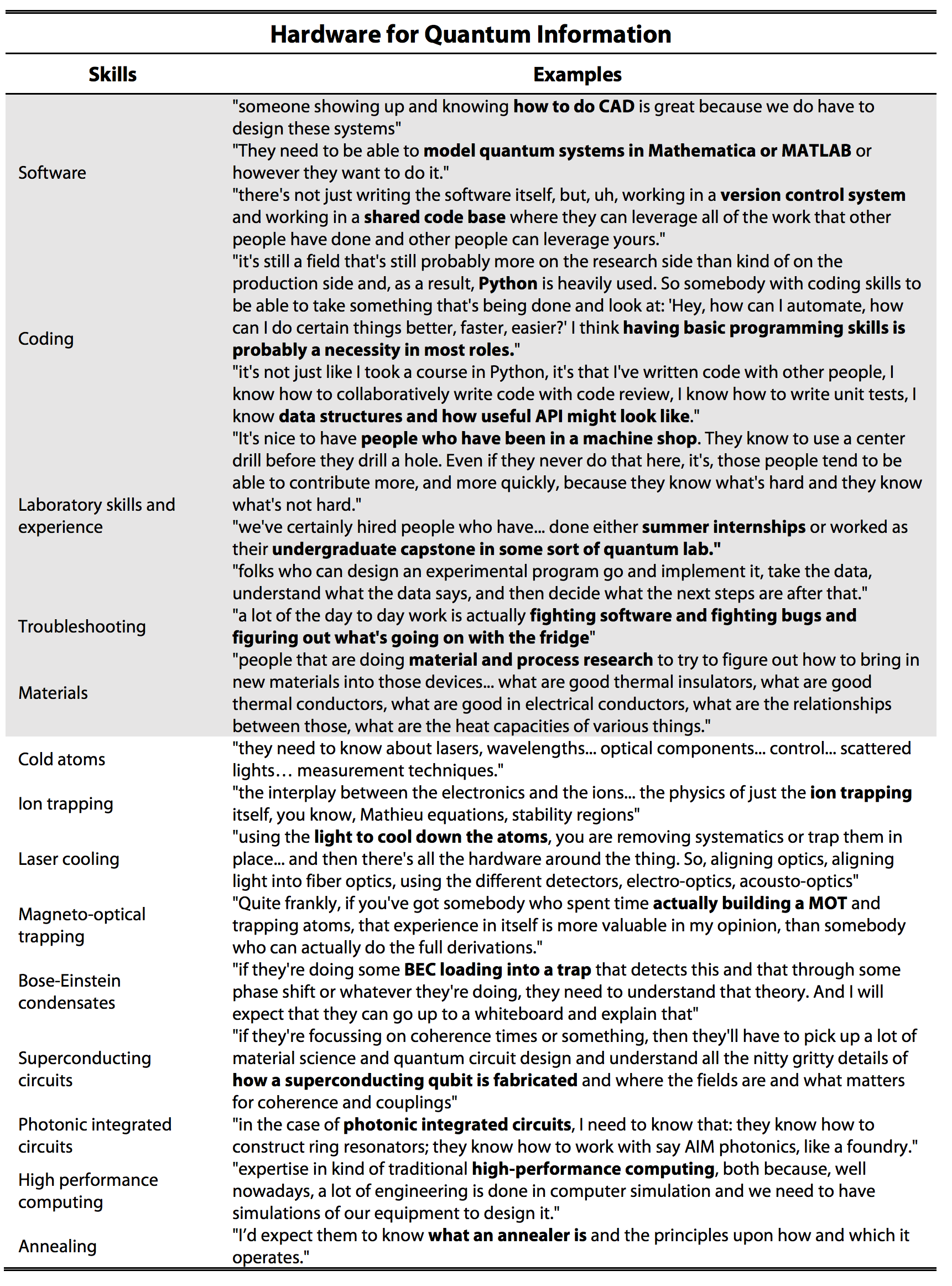}
    \label{tab:hardware}
\end{table}
\begin{table}[h]
    \centering
    \caption{Skills and examples of relevance to the quantum industry for a possible electronics course.}
\includegraphics[width=0.9\textwidth]{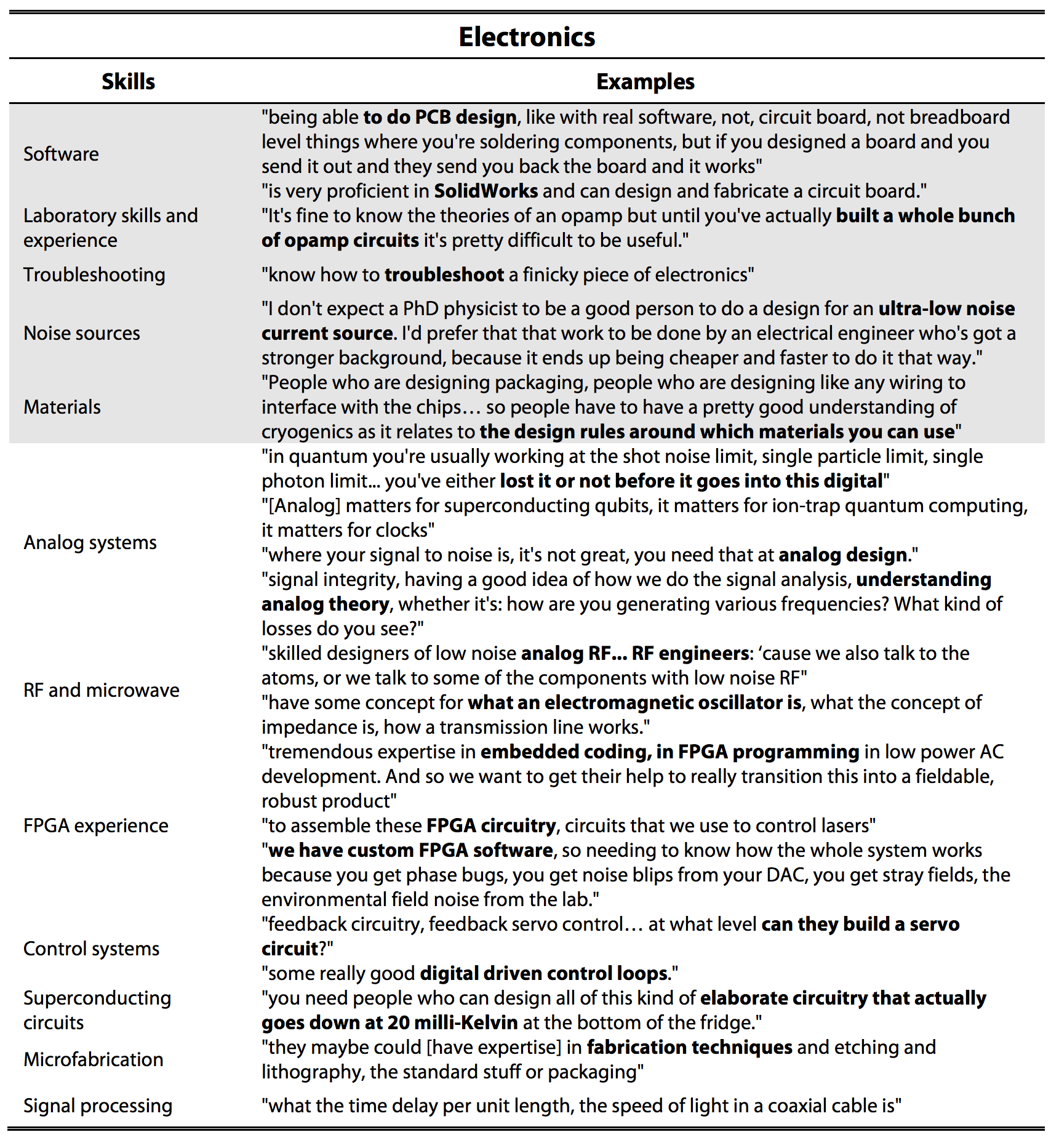}
    \label{tab:electronics}
\end{table}
\begin{table}[h]
    \centering
    \caption{Skills and examples of relevance to the quantum industry for a possible mechanical engineering course.}
\includegraphics[width=0.9\textwidth]{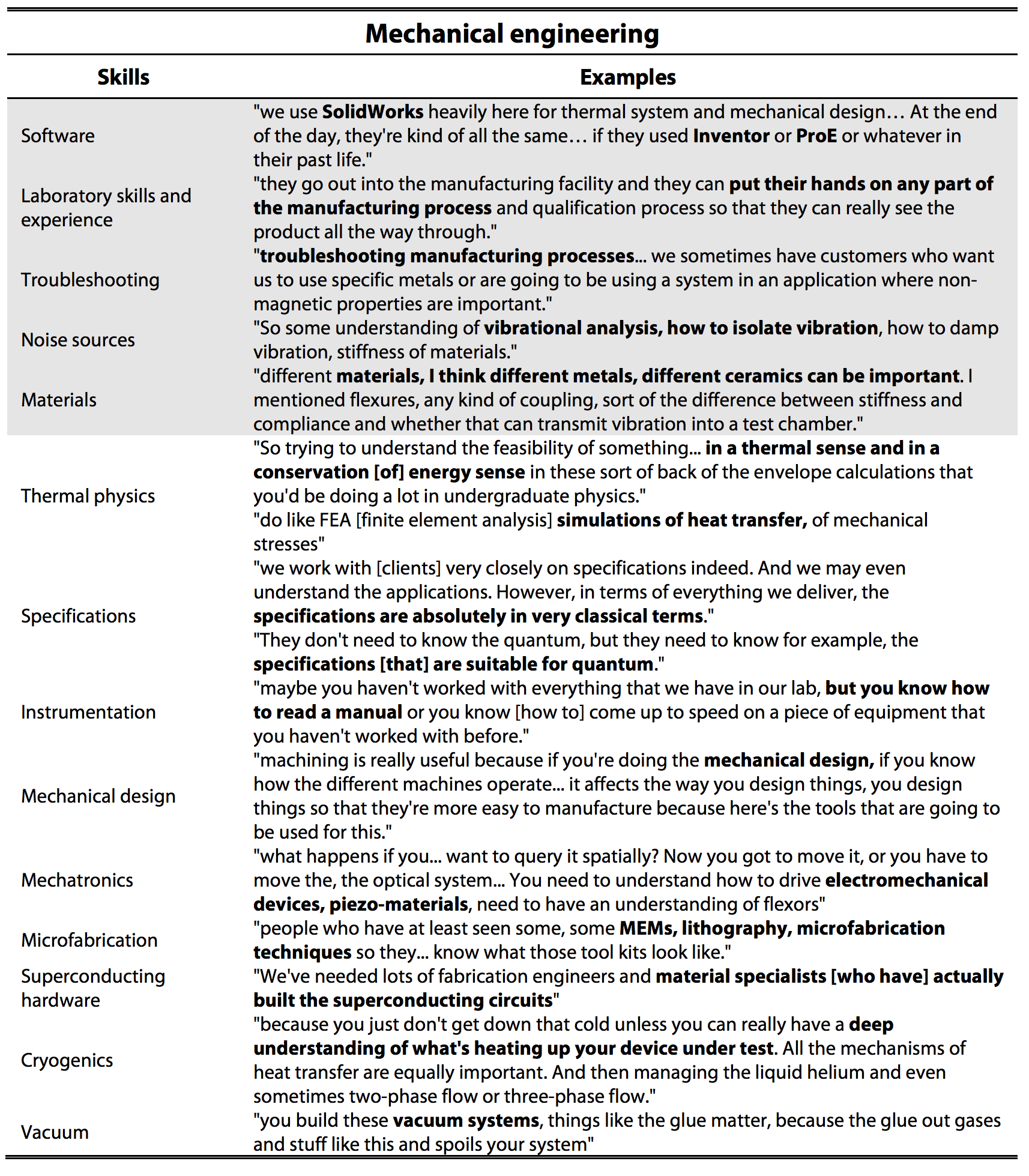}
    \label{tab:mech-eng}
\end{table}
\begin{table}[h]
    \centering
    \caption{Skills and examples of relevance to the quantum industry for a possible optics and opto-mechanics course.}
\includegraphics[width=0.9\textwidth]{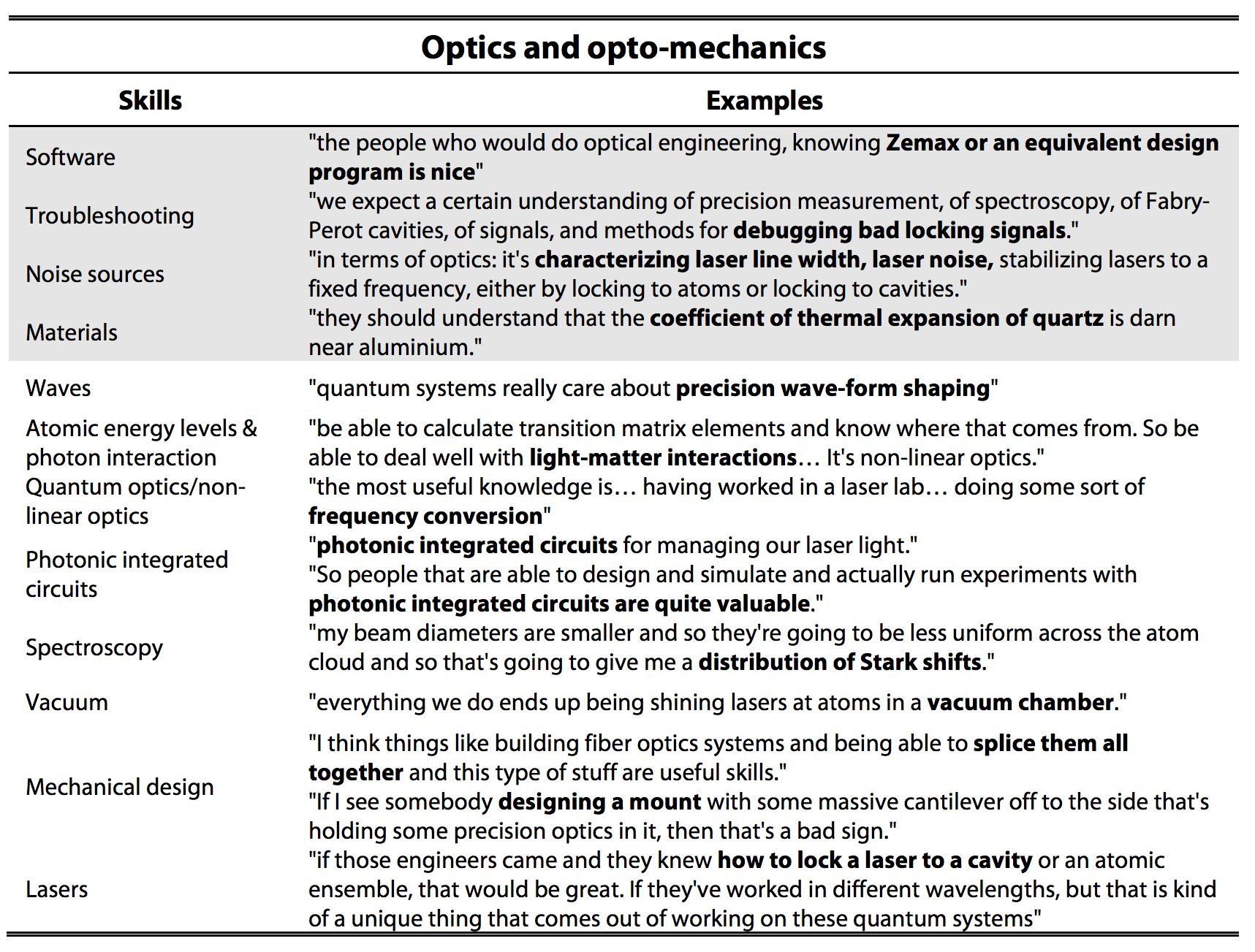}
    \label{tab:optics}
\end{table}

\end{document}